\documentstyle [11pt,epsf] {article}

\begin{document}

\begin{titlepage}

\begin{center}
\vspace{2cm} 
\LARGE The Environmental Dependence of the Relations
between Stellar Mass, Structure, Star Formation and Nuclear Activity
in Galaxies \\
\vspace{1cm} 
\large
Guinevere Kauffmann$^1$, Simon D.M. White$^1$, Timothy M. Heckman$^2$, Brice M\'enard$^{3,1}$,   
Jarle Brinchmann$^{1,4}$, St\'ephane Charlot$^{1,5}$, Christy Tremonti$^6$, Jon Brinkmann$^7$\\
\vspace{0.3cm}
\small
{\em $^1$Max-Planck Institut f\"{u}r Astrophysik, D-85748 Garching, Germany} \\
{\em $^2$Department of Physics and Astronomy, Johns Hopkins University, Baltimore, MD 21218}\\
{\em $^3$Institute for Advanced Study, Einstein Drive, Princeton, NJ 08540}\\
{\em $^4$ Centro de Astrofisica da Universidade do Porto,
Rua das Estrelas, 4150-762 Porto, Portugal}\\
{\em $^5$ Institut d'Astrophysique du CNRS, 98 bis Boulevard Arago, F-75014 Paris, France} \\
{\em $^6$ Steward Observatory, 933 North Cherry Ave, Tuscon, AZ 85721} \\
{\em $^7$ Apache Point Observatory, P.O. Box 59, Sunspot, NM 88349} \\
\vspace{0.6cm}
\end{center}
\normalsize
\begin {abstract}
We use a complete sample of galaxies drawn from the Sloan Digital Sky
Survey (SDSS) to study how structure, star formation and nuclear
activity depend on local density and on stellar mass. Local density is
estimated by counting galaxies above a fixed absolute magnitude limit
within cylinders 2 Mpc in projected radius and $\pm$500 km s$^{-1}$ in
depth.  The stellar mass distribution of galaxies shifts by almost a factor
of two  towards higher masses between low and high density
regions. At fixed stellar mass both star formation and nuclear
activity depend strongly on local density, while structural parameters such as size and
concentration are almost independent of it. Only for low mass                
galaxies ($M_* < 3 \times 10^{10} M_{\odot}$) do we find a weak   
shift towards greater concentration and compactness in the highest density regions.
The galaxy property most
sensitive to environment is specific star formation rate. For galaxies
with stellar masses in the range $10^{10} - 3\times 10^{10}
M_{\odot}$, the median SFR/M$_*$ decreases by more than a factor of 10
as the population shifts from predominantly star-forming at low
densities to predominantly inactive at high densities.  This decrease 
is less marked, but still significant  for high mass galaxies. 
Galaxy properties that are associated with star formation correlate
strongly with local density.  At fixed stellar mass twice as many
galaxies host AGN with strong [OIII] emission in low density regions
as in high. Massive galaxies in low-density environments also contain
more dust.  To gain insight into the processes that shut down star
formation, we analyze correlations between spectroscopic indicators
that probe the SFH on different timescales: the 4000 \AA\ break
strength, the Balmer-absorption index H$\delta_A$, and the specific
star formation rate SFR/$M_*$. The correlations between these
indicators do not depend on environment, suggesting that the decrease
in star formation activity in dense environments occurs over long
($>1$ Gyr) timescales. 
Since structure does not depend on environment for galaxies with 
masses greater than $3 \times 10^{10} M_{\odot}$,  the trends
in recent star formation history (SFH), dust and nuclear activity in
these systems cannot be  driven  by
processes that alter structure, for example mergers or harrassment. 
The SFH-density correlation is strongest for
small scale estimates of local density. We see no evidence that star
formation history depends on environment more than 1 Mpc from a
galaxy.  Finally, we highlight a striking similarity between the
changes in the galaxy population as a function of density and as a function
of redshift. We use mock catalogues derived from N-body simulations to
explain how this may be understood.
\end {abstract}
\vspace {0.8 cm}
Keywords: galaxies:formation,evolution; 
galaxies: stellar content 
\end {titlepage}
\normalsize

\section {Introduction}

One of the most fundamental correlations between the properties of
galaxies in the local Universe is the so-called morphology-density
relation. This relation was first quantified by Oemler (1974) and
Dressler (1980) who showed that star-forming, disk-dominated galaxies
reside in lower density regions of the Universe than inactive
elliptical galaxies.  The standard morphological
classification scheme mixes elements that depend on the structure of a
galaxy (disk-to-bulge ratio, concentration, surface density) with
elements related to its recent star formation history (dust-lanes,
spiral arm strength). It is by no means obvious that these two
elements should depend on environment in the same way.  The physical
orgin of the morphology density relation is still a subject of debate.
Much of the argument centres on whether the relation arises early on
during the formation of the object (the so-called `nature'
hypothesis), or whether it is caused by density-driven evolution
(the `nurture' scenario).

There are many different physical processes that could in principle
play a role in driving the morphology-density relation. Mergers or
tidal interactions can destroy galactic disks and convert spiral and
irregular galaxies into bulge-dominated ellipticals and S0s 
(Toomre \& Toomre 1972; Farouki \& Shapiro 1981).
Mergers operate most efficiently in galaxy groups. In clusters merging
cross-sections are low, but galaxies may still be affected by the
cumulative effect of many weaker encounters
(Richstone 1976;  Moore et al 1996).
Interactions with the dense intracluster gas may also strip away the
interstellar medium of a galaxy and cause a strong reduction in its
star formation rate (Gunn \& Gott 1972).  Finally, gas cooling processes are also strongly
dependent on environment.  In dark matter halos with low masses,
infalling cold gas is never shock-heated and will collapse directly
onto the galaxy. In high mass halos, collapsing gas is first heated to
the halo virial temperature by shocks;  it  then remains
pressure-supported and in quasi-static equilibrium while it cools by
radiative processes (White \& Frenk 1991; Birnboim \& Dekel 2003).

Disentangling the processes responsible for                       
the observed correlations has proved to be a difficult task. However,
the subject has received much  impetus from the completion of  
large spectroscopic and photometric surveys of nearby
galaxies.  There have been many recent papers analyzing trends in
galaxy colours and emission line equivalent widths as a function of
local galaxy density and as a function of distance from the centres of
clusters. These studies indicate that the correlation between
the star formation history of a galaxy and its environment extends to
low densities and to large clustercentric radii (e.g. Kodama et al
2001; Gomez et al 2003; Lewis et al 2002; Pimbblet et al 2002; Balogh
et al 2003).  This suggests that the morphology-density relation is
not driven by processes that operate only in extreme environments,
such as the centres of rich clusters.

There have also been studies of the relative strength with which
different galaxy properties correlate with density. These 
indicate that galaxy colour is the property most tightly linked to    
environment. Galaxy luminosity also depends on environment,
but at fixed luminosity, the structural parameters and surface
brightnesses of galaxies depend only weakly on density (Blanton et al
2003c).  These results suggest that processes such as merging or
harrassment, do not drive the primary                                               
correlation between colour and density. Other recent studies of the
effect of environment on galaxy properties using large galaxy surveys
include the study of the morphology-density relation by Goto et al (2003a)
and the study of the frequency of AGN as a function of environment
by Miller et al (2003).

Over the past few years, we have been developing new methods for
constraining the {\em physical properties} of galaxies using the
wealth of information contained in the high signal-to-noise,
high-resolution galaxy spectra from the Sloan Digital Sky Survey
(hereafter SDSS; York et al 2000; Stoughton et al 2002 Abazaijan et al
2003). We have applied these methods to a sample of $\sim$ 100,000
galaxies in the SDSS Data Release One (DR1).  We used two stellar
absorption line indicators, the 4000 \AA\ break strength D$_n$(4000)
and the Balmer absorption line index H$\delta_A$, to constrain the
mean stellar age and star formation history of each galaxy.  A
comparison with broad band magnitudes then yields estimates of dust
attenuation and of stellar mass. Our methodology for deriving stellar
mass, dust attenuation strength and burst mass fraction is described
in Kauffmann et al (2003a;hereafter Paper I ). We have also used the
emission lines in the spectra to identify galaxies with active nuclei
(AGN).  We then use the comined population synthesis and
photo-ionization models of Charlot \& Longhetti (2001) to derive
gas-phase metallicity, star formation rate, ionization parameter,
dust-to-gas ratio and extinction for emission-line galaxies without
AGN.  Results using physical parameters derived from both absorption
and emission lines have been presented in Kauffmann et al
(2003b,c), Brinchmann et al (2004) and Tremonti et al (2004).

Our studies have revealed that many of the physical properties of
galaxies are very strongly correlated. Almost all galaxy properties
depend strongly on stellar mass.  Massive galaxies have old stellar
ages, high mass-to-light ratios, low star formation rates and little
dust attenuation They also have high concentrations and stellar
surface mass densities and they frequently host AGN.  Low mass
galaxies have young stellar ages, low mass-to-light ratios and are
usually actively forming stars at the present day.  They have low
concentrations and stellar surface mass densities and they almost
never contain active nuclei.  There is also a tight relation between
stellar mass and gas-phase metallicity for the emission line galaxies
in the sample.  Although many of these correlations were already known
from previous analyses, the very large number of galaxies in the SDSS
sample has allowed us to quantify the {\em distribution} of different
galaxy properties as a function of mass. Our analysis has also
revealed that there is a characteristic stellar mass ($\sim 3 \times
10^{10} M_{\odot}$) where many of these properties appear to change
very rapidly.     

In this paper, we attempt to gain more insight into the nature of
these relations by studying how they depend on environment. According
to the standard cosmological paradigm , structure in the present-day
Universe formed through a process of hierarchical clustering, with
small structures merging to form progressively larger ones.  The
theory predicts that density fluctuations on galaxy scales collapsed
earlier in regions that are currently overdense.  Galaxies in high
density regions of the Universe such as galaxy clusters are thus more
`evolved' than galaxies in low density regions or voids.  As we have
discussed, not only did galaxies in dense regions ``form'' earlier,
but they have also been more subject to processes such as stripping
and harrassment that operate preferentially in such environments.  By
studying how the relations between different galaxy properties vary
with density, we hope to deduce whether these relations were
established early on when the galaxy first assembled (nature) or
whether they are the end product of a set of physical processes that
have operated over the whole history of the Universe (nurture).

Our paper is structured as follows: Section 2 reviews the properties
of the galaxy sample. Section 3 discusses our methods for
characterizing the environments of the galaxies in our sample. Section
4 re-introduces the stellar mass partition function of Paper I and
discusses how it varies with environment. In section 5, we discuss how
the correlations between different galaxy properties depend on
density. In section 6, we attempt to place constraints on the
scale-dependence of the environmental effect. Finally in section 7, we
interpret our results in the light of modern theories of structure
formation.

\section {Review of the Spectroscopic Sample of Galaxies}

The on-going Sloan Digital Sky Survey
is using a  dedicated 2.5-meter wide-field
telescope at the Apache Point Observatory to conduct an imaging and
spectroscopic survey of about a quarter of the sky. The imaging is
conducted in the $u$, $g$, $r$, $i$, and $z$ bands (Gunn et al. 1998;
Hogg et al. 2001; Fukugita et al. 1996; Smith et. al. 2002),
and spectra are obtained with a pair of multi-fiber spectrographs.
When the survey is complete, spectra will have
been obtained for nearly 10$^6$ galaxies and 10$^5$ QSOs selected
from the imaging data.
The results in this paper are   
based on spectra of $\sim$122,000 galaxies with 
$14.5 < r < 17.77$  contained in the the SDSS Data
Release One (DR1). These data were
made publically available in 2003.
Details on the spectroscopic target selection
for the ``main'' galaxy sample  can be found in
Strauss et al. (2002) and description of the tiling algorithm is given in Blanton et al (2002).
The reader is referred to Pier et al (2003) for details about the astrometric calibration.

The spectra are obtained
through 3 arcsec diameter fibers.
At the median redshift, the spectra
cover the rest-frame wavelength range from $\sim$3500 to 8500 \AA\
with a spectral resolution $R \sim$ 2000 ($\sigma$$_{instr} \sim$
65 km/s). The spectra are spectrophotometrically calibrated through
observations of F stars in each 3-degree field.
By design, the spectra are well-suited to the determinations
of the principal properties of the stars and ionized gas in galaxies.
The spectral indicators (primarily the 4000 \AA \hspace{0.1cm} break and
the H$\delta_A$ index) and the emission line fluxes used to
determine the star formation rates  analyzed  in this paper are calculated using a 
special-purpose code described in detail in Tremonti et al (2004, in preparation).
A detailed description of the galaxy sample and the methodology used to  derive
parameters such as stellar mass and dust attenuation strength
can  be found in Paper I. A description of the methods for deriving physical
parameters from the emission line measurements can be found in Brinchmann et al (2003).

The SDSS imaging data provide the basic structural parameters that are used in this analysis.
The $z$-band absolute magnitude, combined with our estimated values of M/L and dust attenuation
$A_z$ yield the stellar mass ($M_*$). The half-light radius in the $z$-band and the
stellar mass yield the effective stellar surface mass-density
($\mu_* = M_*/2\pi r_{50,z}^2$). As a proxy for Hubble type we use
the SDSS ``concentration'' parameter $C$, which is defined as the ratio
of the radii enclosing 90\% and 50\% of the galaxy light in the $r$ band
(see Stoughton et al. 2002). Strateva et al. (2001) find that galaxies
with $C >$ 2.6 are mostly early-type galaxies (E, S0 and Sa), whereas spirals and irregulars
have 2.0 $< C <$ 2.6. Throughout this paper, we have assumed a cosmology
with $\Omega=0.3$, $\Lambda=0.7$ and H$_0$= 70 km s$^{-1}$ Mpc$^{-1}$.

\section {The Definition of the Density Estimator}

Two main approaches have been used  to characterize the environments of galaxies
in redshift surveys. One  is to assign galaxies to groups or to clusters. If a group
contains a sufficient number of galaxies with measured redshifts, its velocity dispersion can be
computed and a  correspondence can be made with a virialized dark matter halo of
some circular velocity or mass. The main drawback of this approach is that there is no  
set procedure for finding groups or clusters. Many different methods have been proposed and      
each must be 
tested and calibrated using mock catalogues drawn from N-body simulations
(Nolthenius \& White 1987; see  Marinoni et al 2002 and
Miller et al 2004  for more recent discussions). 
In addition, only a relatively small fraction of
galaxies in a flux-limited sample can be assigned to a group or cluster.
Galaxies with only a few close neighbours are generally excluded from analysis.

In the second approach, a local density is estimated for each galaxy in the sample. Following
the early work of Dressler (1980), many authors have estimated density using the projected distance to
the $n$th nearest spectroscopically-observed neighbour, with $n$ in the range $3-10$ 
(e.g. Hashimoto et al 1998; Gomez et al 2003; Lewis et al 2003; Tanaka et al 2003).
Because the density of galaxies varies with distance in a magnitude-limited survey, most such    
analyses are restricted to volume-limited sub-samples of the full survey.

Other authors have estimated density using galaxy counts within fixed metric
apertures. For example, recent studies of galaxy properties 
as a function of density in the SDSS survey 
(Blanton et al (2003a,b) and Hogg et al (2003a,b)) have used  a variety of estimators,
including counts of  galaxies in the spectroscopic sample  within spheres or cylinders  8 $h^{-1}$ Mpc
in radius and counts in the imaging data smoothed over  scales of
 1 $h^{-1}$Mpc. 

In this paper, we follow an approach similar to that of Blanton, Hogg and collaborators. 
We calculate local
density using galaxy counts to a fixed absolute magnitude  inside a fixed volume. Our choice of 
``target'' galaxies and of the volume within which we evaluate the counts
around them are motivated by two important considerations:
\begin{enumerate}
\item We wish to study galaxy properties over a large range in stellar mass and hence
over a large range in absolute magnitude. Our
target sample must therefore extend to low redshifts in order to include low mass galaxies.
\item We wish to study not only how galaxy properties such as mass or star 
formation history correlate with density, 
but also how the {\em relations} between these properties change with environment.
We thus require a large  sample of targets.
\end {enumerate}

Together these considerations mandate a density estimator 
that can be applied uniformly over a reasonably large
range in redshift. As discussed by Blanton et al (2003a), signal-to-noise considerations 
favour galaxy counts  calculated  within large volumes. Theoretical considerations,
however, suggest that one should estimate the local density within
a small volume. As shown by Lemson \& Kauffmann (1999),
the only property of a dark matter halo that is strongly 
correlated with environment on large scales is mass.
Properties such as spin parameter, concentration and formation time show almost no dependence
on the {\em surrounding} dark matter density or on the external tidal field.  
The formation path and the merging history of a dark matter halo
(and so presumably of the galaxies within it) depends on its present-day virial mass
but {\em not} on its larger scale environment. 
Fluctuations on galaxy-scales will
collapse earlier in those regions of the Universe that end up
inside  massive halos at $z=0$ (e.g. Bower 1991; Lacey \& Cole 1993).   
From a theoretical standpoint, it is thus natural
to expect that galaxy properties should  correlate most strongly with densities
evaluated on scales comparable to the virial radii of typical 
halos at the present day ($\sim$ 1 h$^{-1}$ Mpc).

The sample of target galaxies used in this paper was selected to have apparent magnitudes in
the range $14.5 < r < 17.77$ and redshifts in the range $0.03 < z < 0.1$. Galaxies down to a limiting
$r$-band magnitude of $-18.1$ are thus included in the sample. In Paper I we                 
showed that the maximum $r$-band stellar mass-to-light ratio for galaxies 
of this absolute magnitude is $\sim 1.5 $.
This means that our sample should be complete down to stellar masses
of $2 \times 10^9$ M$_{\odot}$.

We have constructed a volume-limited sample of ``tracer'' galaxies that
we use to evaluate the density around every target. We have used the
photometric galaxies in DR1 that overlap with the spectroscopic survey,
namely the stripes 9-12, 33-36, 76, 82 and 86. In practice, we first
select target galaxies that lie sufficiently far away from the nearest
survey boundary or missing fields. This reduces the number of usable
target objects, but it allows us to work with a density estimator that is
well defined for all target galaxies and that does not suffer from
problems due to edge effects.
We then count the number of spectroscopically-observed galaxies
that are located within 2 Mpc in projected radius and $\pm 500$ km s$^{-1}$ in velocity
difference from the target. We eliminate all galaxies that are predicted to be  fainter than r=17.77
at $z=0.1$ (we adopt the K-corrections of Blanton et al (2002c)). We also correct for 
galaxies that were  missed as a result of fibre collisions. This is done by counting the number
of galaxies in the imaging survey with $14.5 < r < 17.77$ and $R_{proj} < 2$ Mpc ($N_{image}$).
We compare this with the number of galaxies that were targeted spectroscopically ($N_{spectro}$).
We then multiply the number counts within our adopted volume by the correction
factor $f_{corr}= N_{image}/N_{spectro}$ . These correction factors are generally
small ($\sim$ 10-20\%).

Our final sample consists of 46,892 target galaxies. 
The distribution of counts around our sample of targets is shown in Fig. 1.
We plot the cumulative fraction of galaxies with counts less than a given value. Fig. 1
shows that only 20\% of the galaxies in our sample have no neighbours. Around half have 2 or 
more neighbours and 10\% have more than 10 neighbours. The dotted line shows the count distribution
around randomly chosen points within our survey area. This gives some indication of
the expected level of contamination due to chance projections. As can be seen, 76\% of the
randomly placed galaxies have no neighbours and only a tiny fraction have more than a
few. 

As a final comment, we emphasize that our choice of density estimator is a compromise
based on considerations of sample size, signal-to-noise and smoothing scale.                                 
In section 6, we will restrict our analysis to low redshift galaxies
and  explore in more detail the effect of changing the volume within which the density is   
computed.

\begin{figure}
\centerline{
\epsfxsize=8.5cm \epsfbox{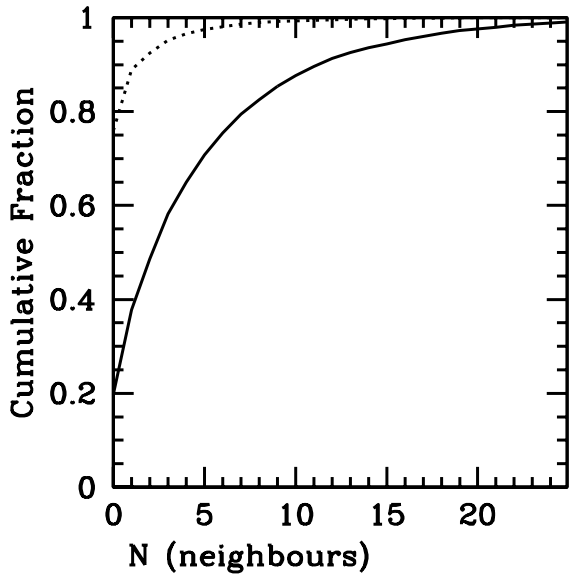}
}
\caption{\label{fig1}
\small
The cumulative fraction of galaxies with counts less than $N$ for the target sample
(solid) and for a random sample (dotted).}
\end {figure}
\normalsize

\section {Environmental Dependence of the Stellar Mass Partition Function}

The study of how the clustering properties of galaxies vary as a function of luminosity,
colour, spectral type and morphology has a long history and a number of different, but
complementary approaches have been used in analyzing the available data.

One approach is to study how the auto-correlation function changes if galaxies
are selected according to different criteria. The early work
of Davis \& Geller (1976) already showed that the two point function
is a strong function of galaxy morphology, being steeper and having a higher amplitude on 
small scales for early-type galaxies than for late-types. 
There now appears to be a consensus
that the {\em slope} of the two-point function is a strong function of the colour
or the spectral type of a galaxy, with red, early-type galaxies exhibiting steeper
slopes than blue, late-type galaxies 
(Willmer, Da Costa \& Pellegrini 1998; Zehavi et al 2002; Madgwick et al 2003; Budavari
et al 2003). If galaxies are split according to luminosity, then the correlation
{\em length} increases for brighter galaxies, but the slope remains approximately
constant (Loveday et al 1995; Norberg et al 2001; Zehavi et al 2002).

There have also been a number of studies addressing how the galaxy luminosity function
changes as a function of environment.
Most authors seem to agree that the characteristic magnitude $M_*$
does vary weakly  with density. It appears to be somewhat brighter in clusters than in the general
field (Balogh et al 2001; De Propris et al 2003) and to be 
fainter in voids (Hoyle et al 2003). This is in qualitative agreement with
correlation function studies described above.
There is, however, no clear  consensus on whether the 
faint-end slope of the luminosity function
depends on environment.

Finally, there have been many recent papers analyzing how galaxy properties
such as colours, luminosities, structural parameters and star formation
rates correlate with local density (e.g. 
Blanton et al 2003a,b; Hogg et al a,b; Gomez et al 2003; Balogh et al 2003; Tanaka et al 2003).
Once again there is general consensus that luminous, red, non-star-forming galaxies
inhabit denser regions of the Universe than faint, blue, star-forming galaxies.
In their most recent paper, Blanton et al (2003b) find that galaxy colour is the
property that is most tightly related to local environment. They also find that
the structural properties of galaxies are less closely related to environment
than their masses and star formation histories.

In this section, we adopt yet another approach. In Paper I, we computed how different kinds of
galaxies  contribute to the total stellar mass budget of the local Universe. Each galaxy in our
sample was given a weight equal to the inverse of the volume
in which it could have been detected. We then
calculated the {\em fraction} of the total stellar mass contained in galaxies as a
function of their stellar mass, D$_n$(4000), $g-r$ colour, size, concentration index and surface
mass density. We found that most of the stellar mass in the nearby Universe 
resides in galaxies
that have stellar masses $\sim 5 \times 10^{10} M_{\odot}$, half-light radii $\sim 3$ kpc and
half-light surface mass densities $\sim 10^9 M_{\odot}$ kpc$^{-2}$.
The distribution of stellar mass as a function of D$_n$(4000) is strongly bimodal,
showing a clear division between galaxies with old stellar populations and galaxies
with more recent star formation.

It is interesting to investigate how the stellar mass partition functions presented
in Paper I vary with environment. This is shown in Fig. 2. We have divided our sample
into 5 density bins and the partition function for each bin is represented by 
a curve with a different colour. Cyan is for galaxies with 0 or 1 neighbour, blue is 
for $N_{neighb}=2-3$,
green is for $N_{neighb}=4-6$, black is for $N_{neighb}=7-11$, and red is 
for $N_{neighb}>12$.

The first three panels of Fig. 2 show how the division of stellar mass among galaxies of
different masses, concentrations and surface densities changes as a function of
density. We find that galaxies in lower density environments 
tend to be less massive, less concentrated
and less dense, but the trends are relatively weak. In contrast, the change in the        
distribution of stellar mass as a function of D$_n$(4000) or $g-r$ colour is quite dramatic.
In the D$_n$(4000) panel, the stellar mass distribution shifts systematically from
the ``blue'' to the ``red'' peak as density increases. 

We also illustrate how the stellar mass is partitioned as a function of the 
normalized star formation rate 
estimated by Brinchmann et al (2003), which is  defined as the present-day star
formation rate (SFR) of the galaxy divided by its stellar mass $M_*$. In
calculating this ratio for Fig. 2, both
the SFR and the stellar mass are estimated for the light within the fibre aperture
(see Brinchmann et al for more details). The distribution of stellar mass
as a function of SFR/M$_*$ shows the same bimodality  as the distribution as a function
of D$_n$(4000). In low-density regions, more than half the stellar mass resides in galaxies
that are presently  forming stars at a rate comparable to their past-averaged one. 
In high-density regions, almost all of the stellar mass resides in galaxies with very
low present-day rates of star formation. (Note that the strong peak at
$\log$ SFR/M$_* =-11.6$ is artificial; all galaxies with very weak or undetectable
star formation have been assigned values in this range).

\begin{figure}
\centerline{
\epsfxsize=14cm \epsfysize=14cm  \epsfbox{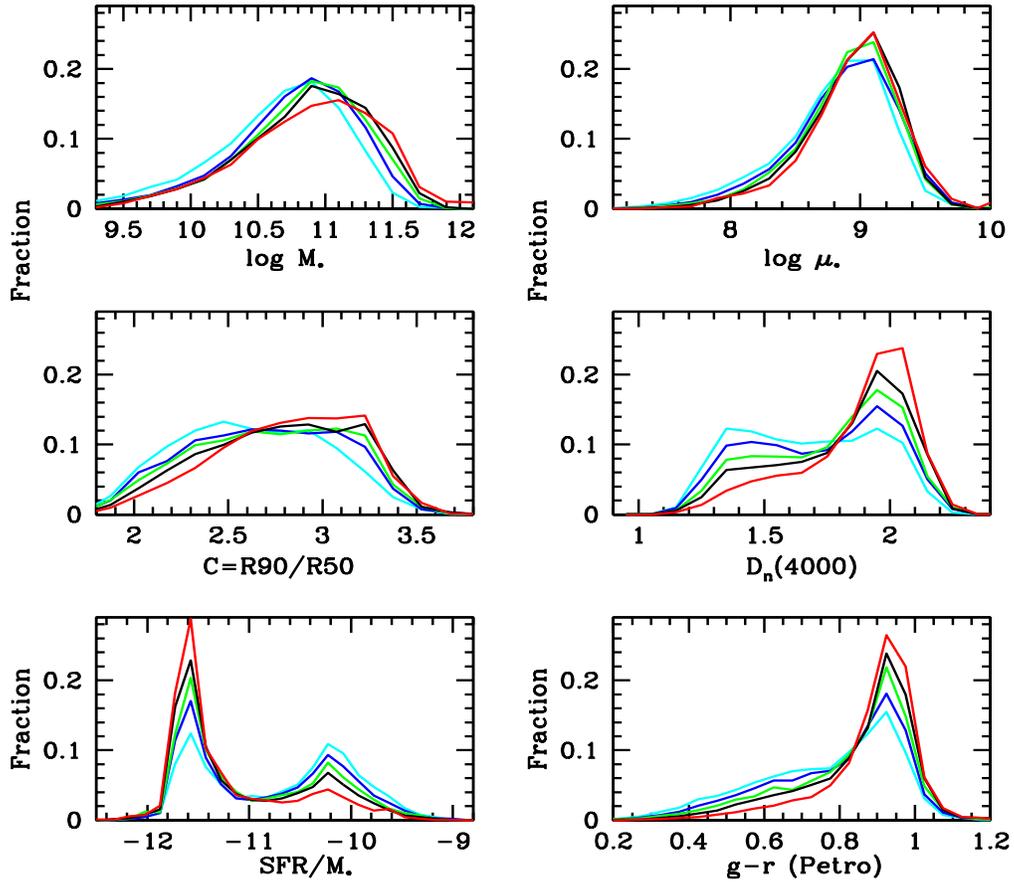}
}
\caption{\label{fig2}
\small
The  fraction of the total stellar mass in the local Universe contained in galaxies
as a function of log M$_*$, log $\mu_*$, concentration index, D$_n$(4000),
SFR/$M_*$ and $g-r$ colour (k-corrected to z=0.1). 
The different colour lines represent the different
density bins as follows: cyan, 0 or 1 neighbour; blue, 2-3 neighbours; green,
4-6 neighbours; black, 7-11 neighbours; red, more than 12 neighbours.}
\end {figure}
\normalsize

\begin{figure}
\centerline{
\epsfxsize=14cm \epsfysize=14cm \epsfbox{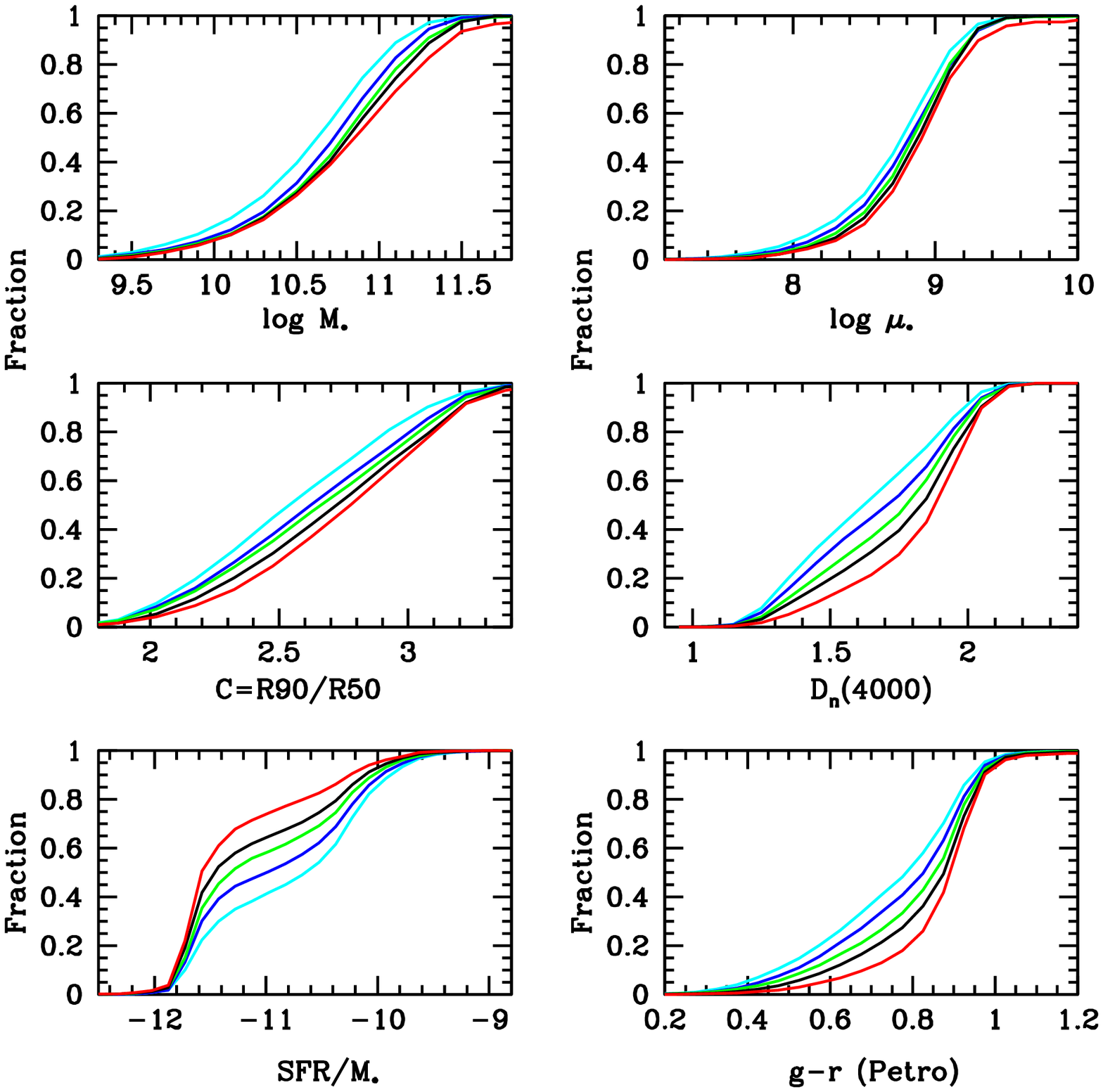}
}
\caption{\label{fig3}
\small
The cumulative fraction of the total stellar mass in the local 
 Universe contained in galaxies
as a function of log M$_*$, log $\mu_*$, concentration index, D$_n$(4000),
SFR/$M_*$ and $g-r$ colour. The different colour lines represent the different
density bins as in Fig. 2.}
\end {figure}
\normalsize

In Fig. 3 we replot the distributions of Fig. 2 in cumulative form. 
Let us define $p(0.5)$ as the value of the
parameter $p$ at a mass fraction of one half. 
 Fig. 2 shows that $M_*(0.5)$ shifts by just under 0.3 dex
from our lowest density bin to our highest density bin. The shift is even smaller
for the stellar  surface mass density $\mu_*$. By contrast, SFR/M$_*$(0.5)
shifts by more than  an order of magnitude from low density to high density regions (the
median value is actually too low to be measured in our highest density bins).
It also useful to transform  the shift into dimensionless units in order
to compare results for different parameters in a uniform way.         
We define a shift $\Delta p$ as
\begin {equation} \Delta p = \frac { \bar{p}_{red} -\bar{p}_{cyan}} {p_{all}(0.95)-p_{all}(0.05)},
\end {equation} 
where $\bar{p}$ is the mass-weighted mean value of parameter $p$
for galaxies in a given density bin, and
$p_{all}(0.95)$ and $p_{all}(0.05)$ are the values of $p$ at mass fractions of 0.95 and 0.05 for
the sample as a whole. Values of $\Delta p$ for different galaxy parameters are listed
in Table 1. As can be seen, the shifts are smallest for the structural parameters $\mu_*$ and $C$ 
and largest for our star formation history indicators  $g-r$, D$_n$(4000) and SFR/M$_*$ .   
We thus conclude, in agreement with Blanton et al (2003b), that  star-formation
history is the galaxy property that is most strongly dependent on environment.   

It is curious that the stellar mass partition function exhibits a clear bimodality
when plotted as a function of D$_n$(4000) or SFR/$M_*$, but not when plotted as 
a function of $g-r$. The colours of star-forming galaxies appear to be more
broadly distributed than their 4000 \AA\ break strengths. This is because the colours
of star-forming galaxies are significantly affected by dust attenuation. 
In Fig. 4 we show how stellar mass divides among galaxies with different
amounts of attenuation. The $z$-band  attenuation A$_z$ is estimated by comparing 
stellar absorption line indicators with galaxy colours, as described in Paper I.
As can be seen, galaxies in low density regions of the Universe typically contain
more dust than galaxies in high density regions.
This is not unexpected, because we have already shown that a larger fraction
of the stellar mass in low-density environments  is in  star-forming
and hence gas-rich galaxies.

\pagebreak
\normalsize

{\bf Table 1:} Relative shift $\Delta p$  in the mass-weighted mean value of
the parameter from the lowest density (cyan) to the highest density bin (red).  
\vspace {0.3cm}

\begin {tabular} {lc}
Parameter  &  $\Delta p$\\                                   
log $M_*$  & 0.164  \\
log $\mu_*$  & 0.126 \\
C  & 0.124 \\
D$_n$(4000) & 0.252 \\
SFR/$M_*$ & 0.271 \\
$g-r$ & 0.179 \\
A$_z$ & 0.150 \\
\end {tabular}

\begin{figure}
\centerline{
\epsfxsize=8cm \epsfysize=7cm \epsfbox{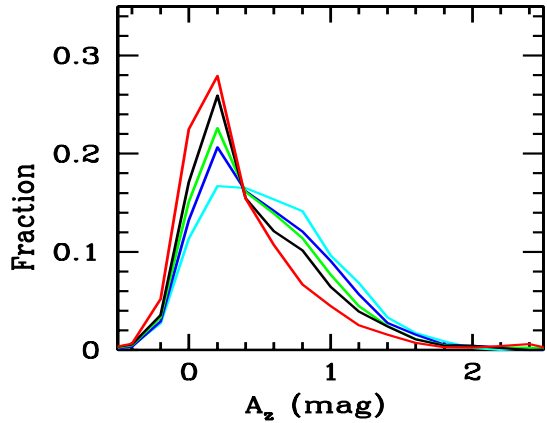}
}
\caption{\label{fig4}
\small
The  fraction of the total stellar mass in the Universe contained in galaxies
as a function of A$_z$, the z-band attenuation in magnitudes. The different 
colour lines represent different local densities as in Fig. 2.}                      
\end {figure}
\normalsize

\section {Environmental Dependence of the Relations between the Physical Properties
of Galaxies}

In Kauffmann et al (2003b; hereafter Paper II), we studied the relations between the
stellar masses, sizes, internal structure, and star formation histories of galaxies.
We showed that strong correlations exist between these different properties.
We also showed that the galaxy population as a whole divides into two distinct ``families''.
Below a characteristic stellar mass of $\sim 3 \times 10^{10} M_{\odot}$, galaxies
have low surface densities and concentrations typical of disk systems. 
They also have young stellar populations and a significant fraction have experienced
recent starbursts. At stellar masses above $3 \times 10^{10} M_{\odot}$, galaxies
have high surface densities and concentrations typical of bulges. The majority
have old stellar populations.

Because properties such as stellar age, mass and surface density 
are so strongly correlated, it is important
to understand not only how a single  property correlates
with environment, but also how the {\em relations} between different properties change
as a function of environment. This is the topic of the present section. 

\subsection {The Relations between Structural Parameters and Stellar Mass}

In Paper II we studied how the                          
surface mass density $\mu_*$ and concentration index $C$ of galaxies vary as a function
of their stellar mass $M_*$. We found that surface mass density correlates  tightly
with stellar mass. Below a mass of $3 \times 10^{10} M_{\odot}$, the median surface
mass density scales with mass as $\mu_* \propto M_*^{0.63}$. At larger masses,
the scaling of $\mu_*$ with $M_*$ becomes weaker and $\mu_*$ eventually saturates at
at a value of around $10^9 M_*$ kpc$^{-2}$.  In Fig. 5 we plot the
$\mu_*-M_*$ and $C-M_*$ relations for galaxies split into                     
six density bins \footnote{Note that the $C$ index has not been corrected for
seeing effects. As demonstrated by Goto et al (2003a), the seeing dependence
of $C$ in the SDSS survey is weak for galaxies  in
the redshift interval $0.05 < z < 0.1$, which is very close the range that is adopted here} . 
The solid lines indicate the median value of $\log \mu_*$ or  $C$
as a function of $\log M_*$, while the dotted lines indicate the 10th and 90th
percentiles of the distributions. Note that we weight each galaxy by
$1/V_{max}$ when we calculate these distributions, but
we only plot the relations over the ranges
in stellar mass where we have at least 200 galaxies per bin.  
We find that for galaxies with
$M_* > 3 \times 10^{10} M_{\odot}$, the $\mu_*-M_*$ and $C-M_*$ relations are 
remarkably {\em independent of density}. Galaxies with masses smaller
than $3 \times 10^{10} M_{\odot}$ have slightly higher concentrations
and surface densities in the very densest environments.

Recall that the surface mass density is proportional to the stellar mass divided by the square of
the half-light radius of the galaxy in the $z$-band. The invariance of the $\mu_*-M_*$
relation for massive galaxies thus implies that the size 
distribution of these galaxies  
does not depend on environment. The concentration index has frequently been used a 
proxy for Hubble type. It was shown by Shimasaku et al (2001) and Strateva et al
(2001) that for bright galaxies there is a good correspondence between
$C$ index  and `by-eye' classification into Hubble type, with
$C \sim 2.6$ marking the boundary between `early-type' galaxies (elliptical, S0s and Sas) and 
`late-type' galaxies  (Sb-Irr).
Conventional widom states that galaxy morphology correlates with local density
(Dressler 1980), so at first glance it is  astonishing to find
no significant variation in the $C-M_*$ relation as a function of environment. 

The most likely solution to this paradox is that `by-eye' classification into Hubble type
is dependent not only on the structure of the galaxy, but also on its star formation
rate. This was noted by Koopmann \& Kenney (1998) who showed that in the Virgo cluster, spirals
with reduced global star formation activity are often assigned 
early-type classifications irrespective of their central light concentrations.
As we will show in the next section, the relations between star formation rate
and the structural parameters of galaxies do vary strongly with environment  for
galaxies of all stellar masses.

\begin{figure}
\centerline{
\epsfxsize=14cm \epsfysize=8cm \epsfbox{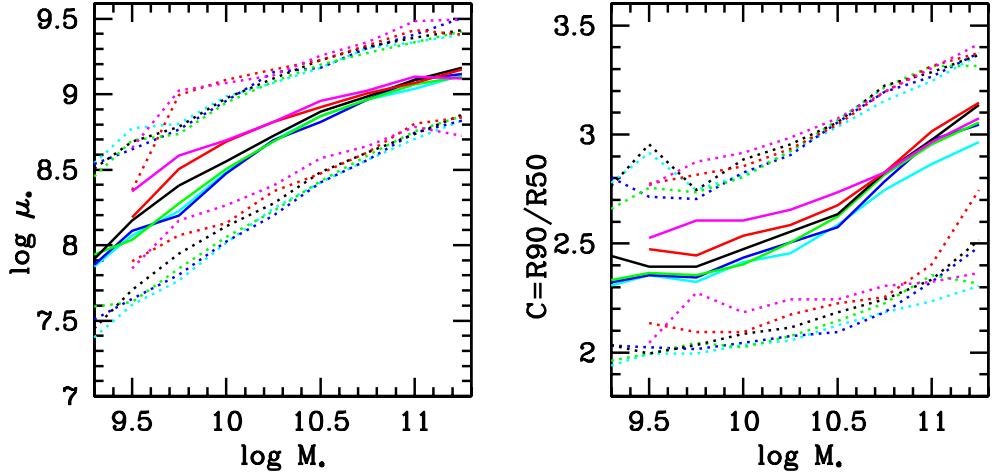}
}
\caption{\label{fig5}
\small
The  relations between surface mass density and stellar mass (left) and concentration
index and stellar mass (right) are plotted for galaxies in six different density
bins as follows: cyan, 0 or 1 neighbour; blue, 2-3 neighbours; green,
4-6 neighbours; black, 7-11 neighbours; red, 12-16 neighbours; magenta,
17 or more neighbours.
The solid curves indicate the median value of $\log \mu_*$ or $C$ and a given
value of $\log M_*$. The dotted lines indicate the 10th and 90th percentiles
of the distributions.}
\end {figure}
\normalsize

\subsection {The Relations between Star Formation History, Stellar Mass and Galaxy Structure} 

\begin{figure}
\centerline{
\epsfxsize=14cm \epsfysize=8cm \epsfbox{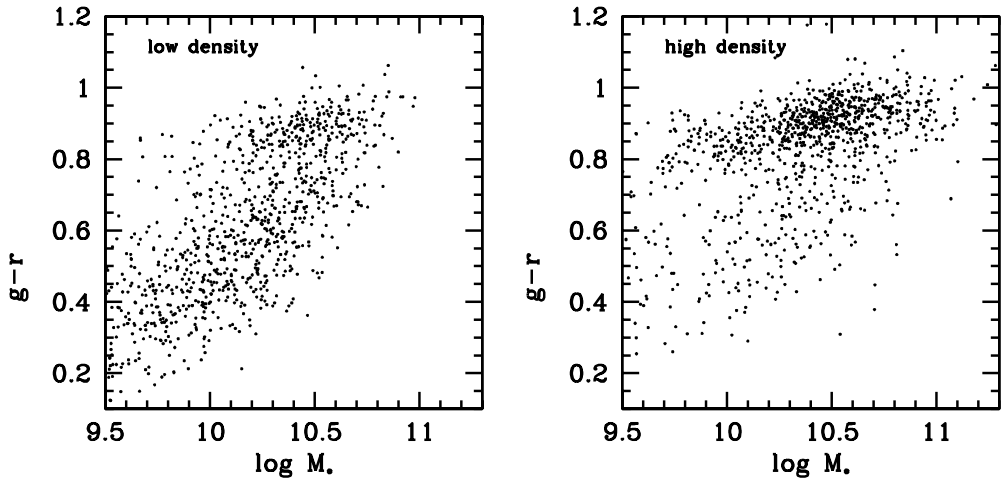}
}
\caption{\label{fig6}
\small
A scatterplot of $g-r$ colour versus stellar mass is plotted for 1000 galaxies in     
our lowest density bin (N$_{neighb}$=0-1; left) and our highest 
density bin (N$_{neighb}>17$; right).} 
\end {figure}
\normalsize

In Paper II, we presented the conditional density distributions of two stellar age
indicators, the 4000 \AA\ break strength D$_n$(4000) and the H$\delta_A$
index as functions of stellar mass, stellar surface density and concentration. 
Brinchmann et al (2003) presented the distribution of the normalized
star formation rate (SFR/$M_*$) as functions of these same parameters. Baldry et al (2003)
studied the distribution of galaxy colours as a function of their absolute
magnitudes. All these studies reach very similar conclusions. Faint low mass galaxies
with low concentrations and surface mass
densities have young stellar populations, ongoing star formation and blue colours. 
Bright galaxies with high stellar
masses, high concentrations and high surface densities have old stellar populations,
little ongoing star formation and red colours. A transition from ``young'' to ``old'' 
takes place at a characteristic stellar mass of $3\times 10^{10} M_{\odot}$, a 
stellar surface density of $3 \times 10^{8} M_*$kpc$^{-2}$, a concentration index
of 2.6 and an r-band absolute magnitude of $\sim -20.3$. 
So how do the relations between stellar age, stellar mass and structural parameters depend
on environment? 

In Fig. 6, we show scatterplots of $g-r$ colour
versus stellar mass for  a random  sample of 1000 galaxies in our lowest density
and highest density bins. This plot clearly illustrates the
``bimodal'' colour-magnitude distribution of galaxies described in Baldry et al (2003);
galaxies separate rather cleanly into a red sequence and
a blue sequence. As density increases, we find that the 
relative {\em weight} of the  two populations shifts from
the blue to the red sequence. In the highest density environments, the red
sequence dominates entirely. This is well-known from studies of galaxy populations
in nearby clusters.

In the top two panels of Fig. 7 we show how the {\em median} values of D$_n$(4000)
and SFR/$M_*$ vary as a function of stellar mass for our different density bins.
(Note that in Fig.7 we have applied the appropriate volume-corrections to our sample.)
The median values
of both parameters shift systematically as density increases. The shift
is largest for galaxies with stellar masses just below our characteristic mass of
$3 \times 10^{10} M_{\odot}$ and corresponds to more than a factor of 10 decrease
in star formation rate from our lowest density to our highest density bins.
At very  high stellar masses ($> 10^{11} M_{\odot}$), 
there is little change in the median SFR with density. Likewise for $M_* < 10^{10} M_{\odot}$,
the effect again becomes somewhat weaker.

The bottom two panels show how the D$_n$(4000)-C and D$_n$(4000)-$\mu_*$ relations change as
a function of density. In the highest density regions,  galaxies with low concentrations
characteristic of disk-dominated systems ($C<2.6$)  have considerably older 
stellar populations than galaxies of similar concentrations in low-density regions. 
Van den Bergh (1976)
was the first to remark on a  class of ``anaemic spirals'', which occur most
often in rich clusters and these systems have recently been studied
in considerable detail using SDSS data by Goto et al (2003b).
In Paper II, we demonstrated that stellar age indicators were more closely correlated with
$\mu_*$ than with $M_*$. This conclusion is  borne out once again in Fig. 7.
The relation between D$_n$(4000)  and $\mu_*$ appears  less sensitive
to environment than the relations between D$_n$(4000) and $M_*$ or $C$.

\begin{figure}
\centerline{
\epsfxsize=14cm \epsfysize=14cm \epsfbox{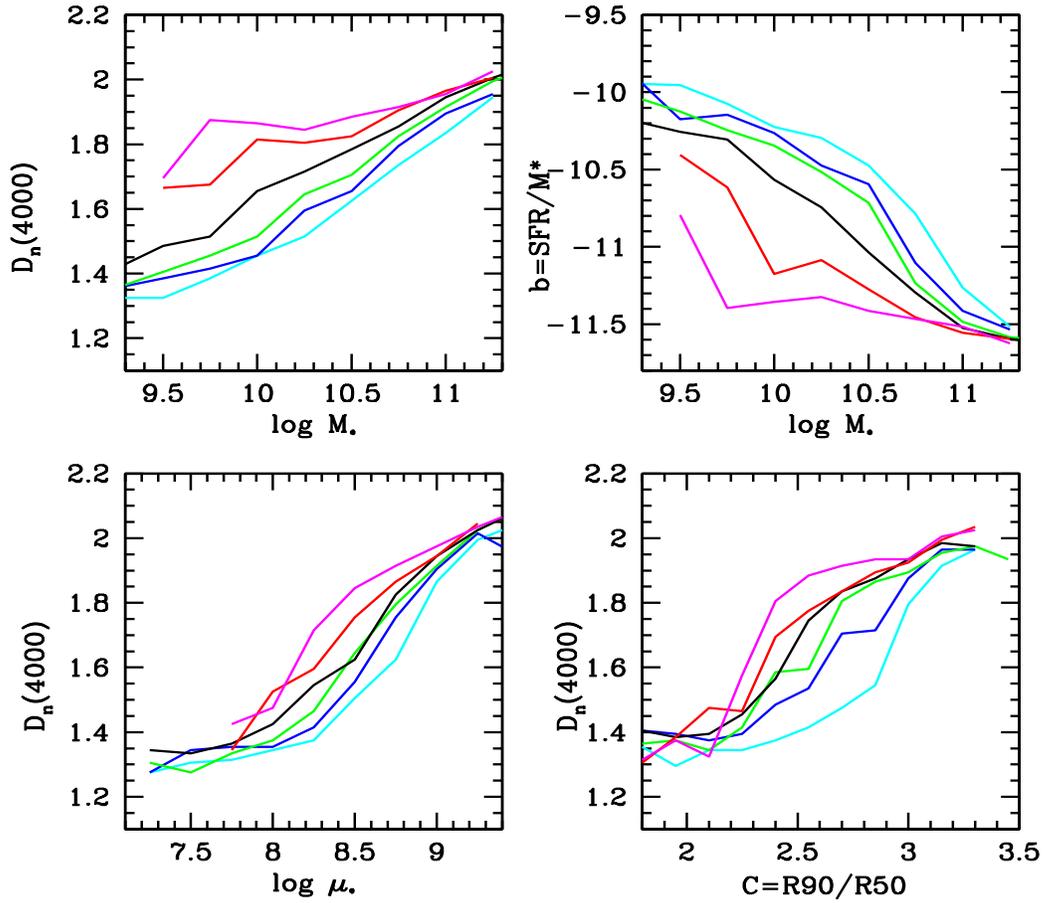}
}
\caption{\label{fig7}
\small
Top: The median relations between D$_n$(4000) and SFR/$M_*$ are plotted as a function of               
stellar mass for 5 different bins in density, colour-coded as in Fig. 5. Bottom: The median relations
between D$_n$(4000) and $\mu_*$ (left) and $C$ (right).} 
\end {figure}
\normalsize
 
\subsection {Relations between Different Indicators of Star Formation History} 

So far we have used the  4000 \AA\ break, the H$\delta_A$ absorption
line index and the normalized star formation rate
(SFR/$M_*$) interchangeably as indicators of the recent star formation history of
a galaxy. In practice, these three indicators probe star formation on rather different timescales:
\begin {enumerate}
\item The normalized star formation rate is estimated using the fluxes of the nebular
emission lines in the galaxy, in particular the H$\alpha$ line (Brinchmann et al 2003).
SFR/$M_*$ thus probes star formation on a timescale
equal to the lifetime of a typical HII region ($\sim 10^7$ years).
\item The strength H$\delta_A$ absorption line peaks once hot O and B stars have
terminated their evolution and the optical light is dominated by late-B to early F-stars.
This occurs about $3\times 10^8$ years after an episode of star formation.
\item The strength of the 4000 \AA\ break increases montotonically with time. At stellar ages
 of more than $\sim 1$ Gyr, metallicity effects also become important.
\end {enumerate}

\begin{figure}
\centerline{
\epsfxsize=14cm \epsfysize=8cm \epsfbox{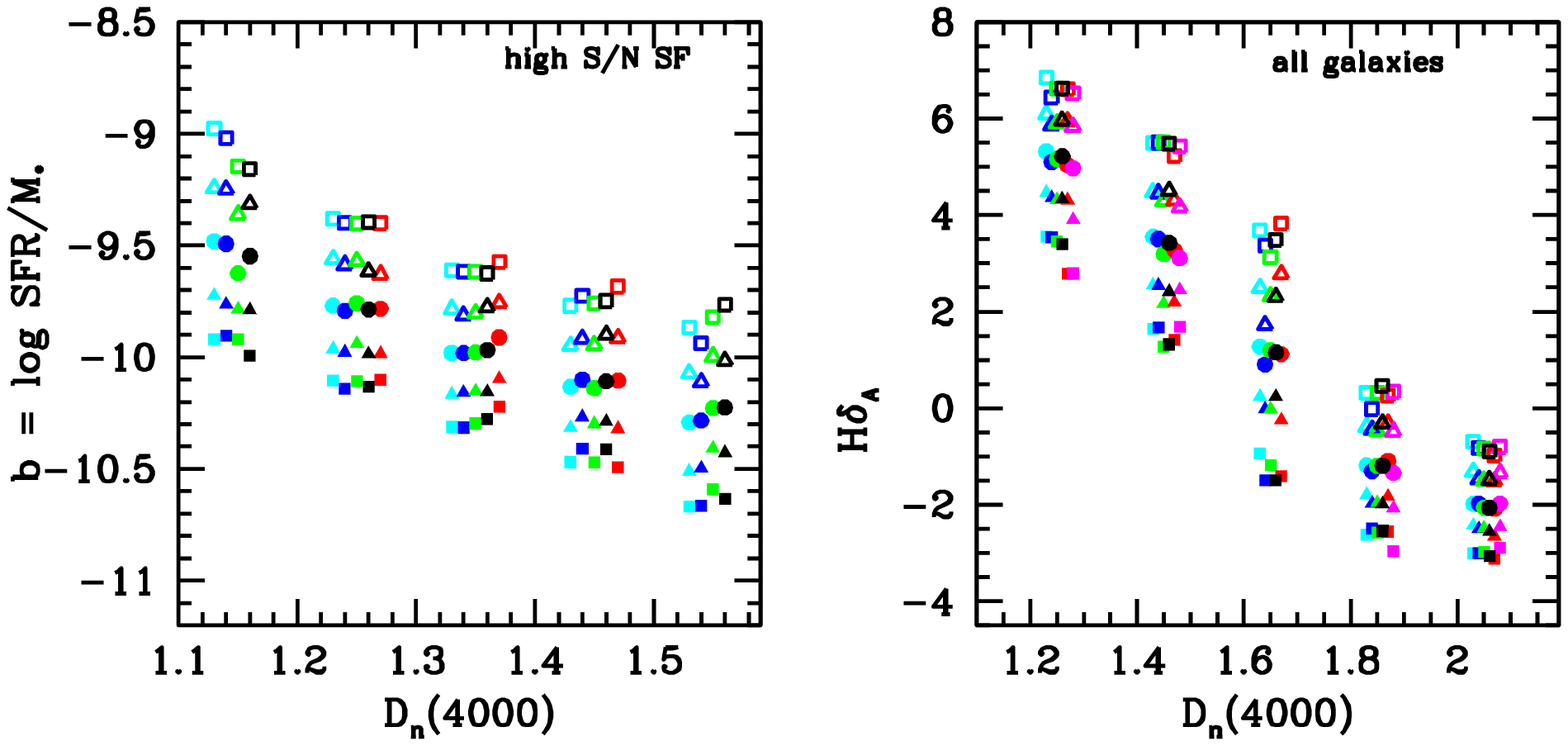}
}
\caption{\label{fig8}
\small
Top: The relations between SFR/$M_*$ and D$_n$(4000) (left) and
H$\delta_A$ and D$_n$(4000) in different density bins colour-coded as
in Figure 5. Solid circles indicate
the median of the distribution, solid and open triangles are 25th and 75th percentiles,
solid and open squares are 10th and 90th percentiles.}              
\end {figure}
\normalsize
 
Insight into the nature of the recent star formation histories of galaxies may be
obtained by studying  correlations between these different indicators.
In Paper II we showed that at a  given value of D$_n$(4000),  
low mass galaxies span a significantly  larger range in H$\delta_A$
equivalent width than high mass galaxies. 
This indicates that the recent star formation histories of low mass galaxies
have been more bursty. This conclusion was confirmed by Brinchmann et al (2003), who
showed that the distribution of SFR/$M_*$ for low mass galaxies exhibited a significant tail         
to high values. This tail was largely absent
for high mass galaxies.

In this section, we investigate whether environment has any detectable influence
on the recent star formation histories of the galaxies in our sample. 
Recent bursts of star formation would
produce a tail of galaxies with large H$\delta_A$ equivalent widths and
large values of SFR/$M_*$ at moderate
values of D$_n$(4000). A recent sudden truncation in star formation (caused by
ram-pressure stripping, for instance) would produce a population of galaxies 
with relatively low 4000 \AA\ break strengths and high H$\delta$  equivalent
widths, but with no detectable emission lines. 

In Fig. 8, we plot the relation between SFR/$M_*$ and D$_n$(4000)  and the relation between
H$\delta_A$ and D$_n$(4000) for galaxies in our different density bins. 
The different symbols indicate different percentiles of the distribution: 
solid and open squares the 10th and 90th percentiles, solid and open triangles the 25th and
75th percentiles, and solid circles  the median. As before, different colours
indicate our different bins in density. Note that we only show the correlation between SFR/$M_*$ 
and D$_n$(4000) for galaxies with $S/N >3$ in the four emission lines
H$\alpha$, H$\beta$, [OIII] and [NII]. For this subset of galaxies, it is possible to
estimate the star formation rate in a reliable way using only    
the emission lines (see Brinchmann et al 2003). Because strong emission-line
galaxies become increasingly rare in high density regions and in galaxies
with old stellar populations, it is not possible
to extend the analysis to the full sample.  The correlation between
H$\delta_A$ and D$_n$(4000), on the other hand, can be  plotted for {\em all} the galaxies in the sample.

As can be seen, the relations between the different indicators show no dependence on
environment. We have also restricted the analysis to galaxies with stellar
masses in the range $3 \times 10^9 -3 \times 10^{10} M_{\odot}$
and we find the same result. In section 5.2 we  demonstrated that galaxies are systematically  
younger in low density regions and older in               
high density regions. Nonetheless, the results in this section
indicate that  the blue galaxies that do exist in high density
regions appear to have recent star formation histories that are indistinguishable
from those in the `field'.
This suggests that the decrease in star formation activity in galaxies
in high density regions  occurs over relatively long ($>$ 1 Gyr) timescales for the
majority of galaxies.
We will come back to this point in section 7.

\subsection {Other Properties associated with Star Formation: AGN and Dust}

\subsubsection {Nuclear activity in galaxies}

In Kauffmann et al (2003c), it was demonstrated that Type 2 (narrow-line) active galactic
nuclei occur almost exclusively in galaxies with stellar masses greater than
$10^{10} M_{\odot}$. Weak AGN (those with [OIII] line luminosities less than 
$10^7$ L$_{\odot}$) occur in massive galaxies with predominantly old stellar
populations, whereas powerful AGN (L[OIII]$> 10^7$ L$_{\odot}$ ) 
are found in massive galaxies containing young stars. Because star formation is
so strongly correlated with environment, it would be natural to expect
that the incidence of powerful AGN would also be a function of local density.

Kauffmann et al (2003c) also  showed that
at luminosities below $10^7 L_{\odot}$, the fraction of detected AGN in SDSS galaxies 
is a strong function of redshift. At larger distances, more
of the host galaxy light falls within the fibre aperture and weak AGN become
progressively more difficult to detect.  At luminosities above
$10^7 L_{\odot}$, there were 
no longer any  distance-dependent selection effects in picking
out AGN using diagnostic diagrams based on emission line ratios.
In the left panel of Fig.9,  we plot the
fraction of galaxies containing strong AGN with L[OIII]$> 10^7 L_{\odot}$ as a 
function of stellar mass in three bins of density. 
As expected, the fraction of
strong AGN in massive galaxies {\em decreases} as a function of density.
Kauffmann et al showed that the fraction of powerful AGN in galaxies with $3 \times 10^{10}-
10^{11} M_{\odot}$ was $\sim 0.1$ for the galaxy population as a whole. 
In our lowest density bin, the fraction rises to $\sim 0.15$ and in our highest density
bin, it is $\sim 0.06$.

\begin{figure}
\centerline{
\epsfxsize=13.5cm  \epsfbox{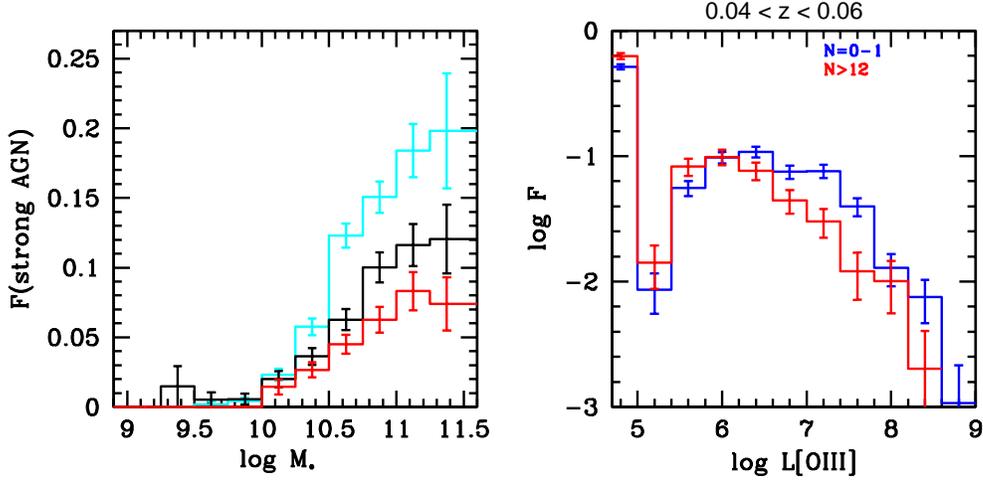}
}
\caption{\label{fig10}
\small
Left: The fraction of galaxies containing AGN with L[OIII]$> 10^7$ $L_{\odot}$
is plotted as a function of stellar mass for  three bins in density. Cyan is for galaxies
with $N_{neighb}=0-1$, black is for $N_{neighb}=7-11$ and red is for $N_{neighb}>12$.
Right: The distribution of [OIII] luminosities 
of AGN in galaxies with stellar masses in the range $3 \times 10^{10} - 3 \times 10^{11} M_{\odot}$
in high density environments (red) and in low density environments (blue).
The sample has been restricted to galaxies with $0.04 < z < 0.06$.
Note that the [OIII] luminosities have been corrected for dust attenuation as described
in Kauffmann et al (2003c).}

\end {figure}
\normalsize

\begin{figure}
\centerline{
\epsfxsize=14cm \epsfysize=8cm \epsfbox{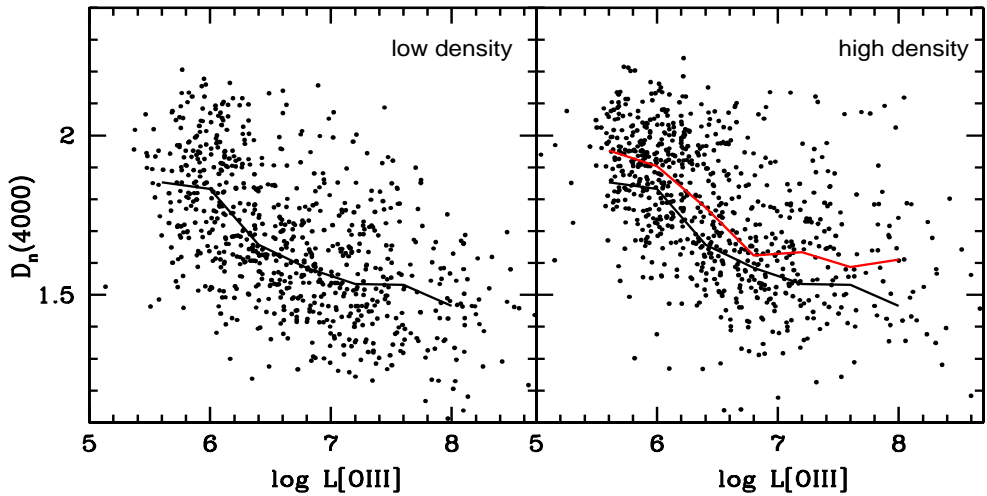}
}
\caption{\label{fig11}
\small
D$_n$(4000) is plotted as a function of L[OIII] for AGN with $0.08 < z < 0.1$
in low density regions (left) and in high-density regions(right).
The black line shows the running median of the distribution
of points in low-density regions and the red line is the same for the
high-density points.}
\end {figure}
\normalsize

\begin{figure}
\centerline{
\epsfxsize=8cm \epsfysize=8cm \epsfbox{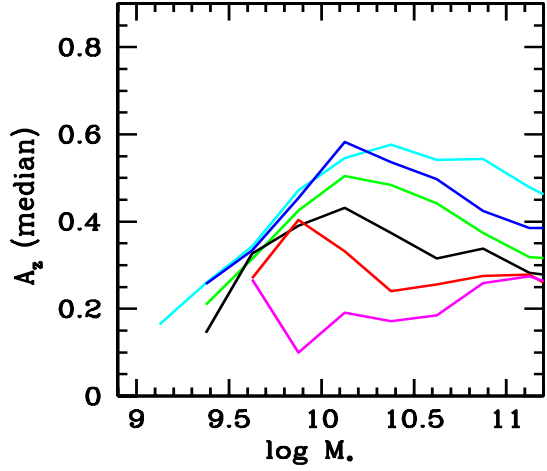}
}
\caption{\label{fig12}
\small
The median $z$-band attenuation
due to dust in magnitudes is plotted as a function of stellar mass in
the same density bins as in Fig. 5}
\end {figure}
\normalsize

\begin{figure}
\centerline{
\epsfxsize=13cm \epsfysize=8cm  \epsfbox{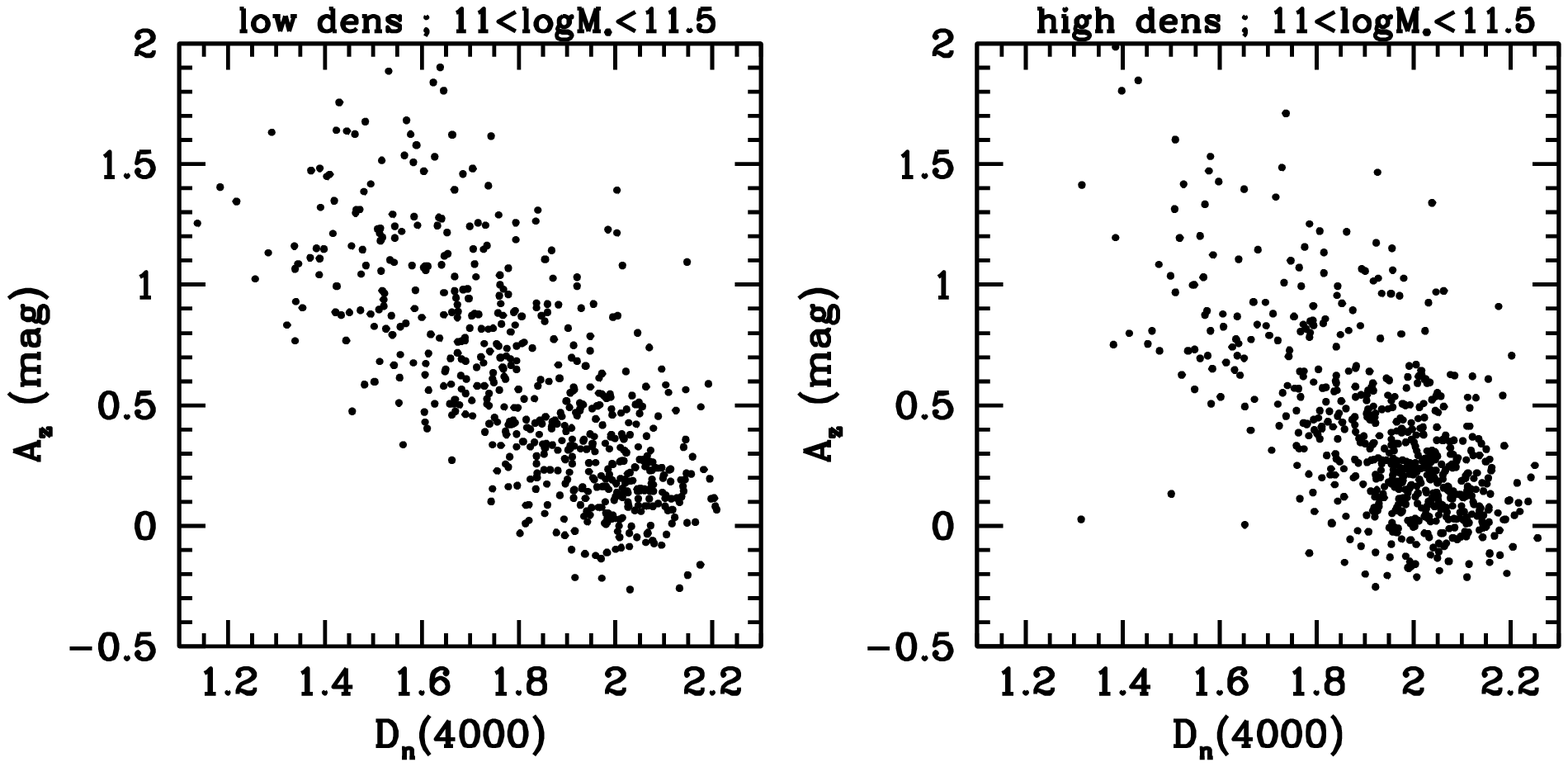}
}
\caption{\label{fig13}
\small
Dust attenuation in the $z$-band  is plotted against D$_n$(4000) for galaxies with 
$11 < \log M_* < 11.5$ in low-density regions (left) and in high-density regions (right).}
\end {figure}
\normalsize

Miller et al (2003) have also used SDSS data to study how the fraction of galaxies
with AGN varies as a function of environment. They found almost no density
dependence and concluded that AGN were unbiased tracers of large-scale structure in the Universe.
Miller et al did not classify their AGN according to [OIII] luminosity,
and it is likely that their sample included a substantial number of weak
AGN. In the right panel of Fig. 9, we plot the distribution of [OIII] luminosities
of AGN in galaxies with stellar masses in the range 
$3 \times 10^{10} -3 \times 10^{11} M_{\odot}$
and in two different bins of density. In order to avoid distance-dependent selection effects,
we have restricted the analysis to galaxies with $0.04 <z < 0.06$. Note that all galaxies
without any detectable AGN are placed into the first bin. As can be seen,
it is the fraction of {\em powerful} AGN with L[OIII]$> 10^7 L_{\odot}$ that depends on
density. As discussed in Kauffmann et al (2003c), AGN with these luminosities are almost all
type 2 Seyfert galaxies.   On the other hand, the fraction of {\em low-luminosity} AGN (LINERs)
depends very little on local density. Since there are many more LINERs than Seyferts,
the overall fraction of detected AGN shows little environmental dependence.

It is interesting to ask whether the physical mechanism responsible for the
decrease in powerful  AGN fraction 
in high density regions is  the same as that responsible for
the decrease in the fraction of star-forming galaxies. If this is the case, then the {\em relation} between
[OIII] luminosity and 4000 \AA\ break strength in AGN should not 
depend on environment.  
In Fig. 10,  we show the relation between D$_n$(4000) and L[OIII]
for AGN in low-density ($N_{neighb} < 2$) and high density
($N_{neighb} > 12$) environments. We find   little change
in the D$_n$(4000)-L[OIII] relation with density. There is a slight shift towards
higher D$_n$(4000) at given L[OIII] in denser regions. 
We have also investigated whether there is any environmental dependence of  the stellar
masses or structural parameters of the host galaxies of AGN of given L[OIII]. 
All these studies indicate that
at  given [OIII] line luminosity, the properties of an AGN host depend very little
on local density. It is the overall {\em fraction} of galaxies that  are able to host
strong-lined AGN that decreases in high-density environments. 

\subsubsection {Dust}

Dust attenuation is another property that can be expected to
be strongly correlated with how much ongoing star formation is taking
place in a galaxy. In Fig. 4 we showed that in low-density regions of the Universe,
a larger fraction of the total stellar mass is in galaxies with significant
dust attenuation.
In Fig. 11, we plot the relation between the median
value of the dust-attenuation in the $z$-band and stellar mass for galaxies in
different bins in density. As can be seen, for galaxies with
$M_* > 10^{10} M_{\odot}$, the amount of dust in a galaxy of given mass
is a relatively strong function of density. For low mass galaxies, the
amount of dust shows little dependence on environment. It is interesting that in low-density 
environments, {\em even very massive galaxies contain a substantial amount of dust}.
Once again it is interesting to ask whether this is simply a consequence of the
fact that there is more star formation in massive galaxies in low-density
environments.

In Fig. 12 we plot
the relation between the 4000 \AA\ break and dust attenuation for massive galaxies
with $10^{11} M_{\odot} < M_* < 3 \times 10^{11} M_{\odot}$ in our lowest density
bin (left) and in our highest density bin (right). As can be seen, the dust
attenuation is tightly correlated with D$_n$(4000).
In low density regions,  a substantial fraction of high mass galaxies have low
4000 \AA\ break strengths and high attenuation. In high density regions, almost
all massive galaxies have both large break strengths and low attenuation. 
However, the  correlation between A$_z$ and D$_n$(4000) is very similar in all
environments. This indicates that the same  mechanism is likely responsible 
for the decrease both in star formation and in dust content for galaxies 
in high density regions of the Universe.

\section {The Scale Dependence of the Environmental Effect}

In the previous two sections, we demonstrated that quantities associated with    
star formation history  -- colour, 4000 \AA\ break, SFR/$M_*$ -- show
strong correlations with local density. In this section we ask whether we can
place any constraints on the {\em physical  scale} over which this correlation occurs.  
As discussed in section 3, 
theoretical considerations lead one to expect that galaxy properties should 
correlate best with dark matter density on 
small ($\sim$ 1 h$^{-1}$ Mpc) scales. There have been some recent
suggestions in the literature, however, that star formation 
shows strong sensitivity to the density of galaxies evaluated on 
scales in excess of 5 Mpc (Balogh et al 2003). 

In order to address this issue, we restrict the analysis in this 
section to galaxies in the redshift
range $0.04 < z < 0.06$. This reduces the total number of galaxies in our sample
by a large factor, but it has the advantage that at low redshifts,
the galaxy population in the SDSS
is densely sampled. It is then possible to explore the effect of an increase or a decrease in 
the size of the volume inside which we evaluate the counts.

\begin{figure}
\centerline{
\epsfxsize=14cm \epsfysize=14cm \epsfbox{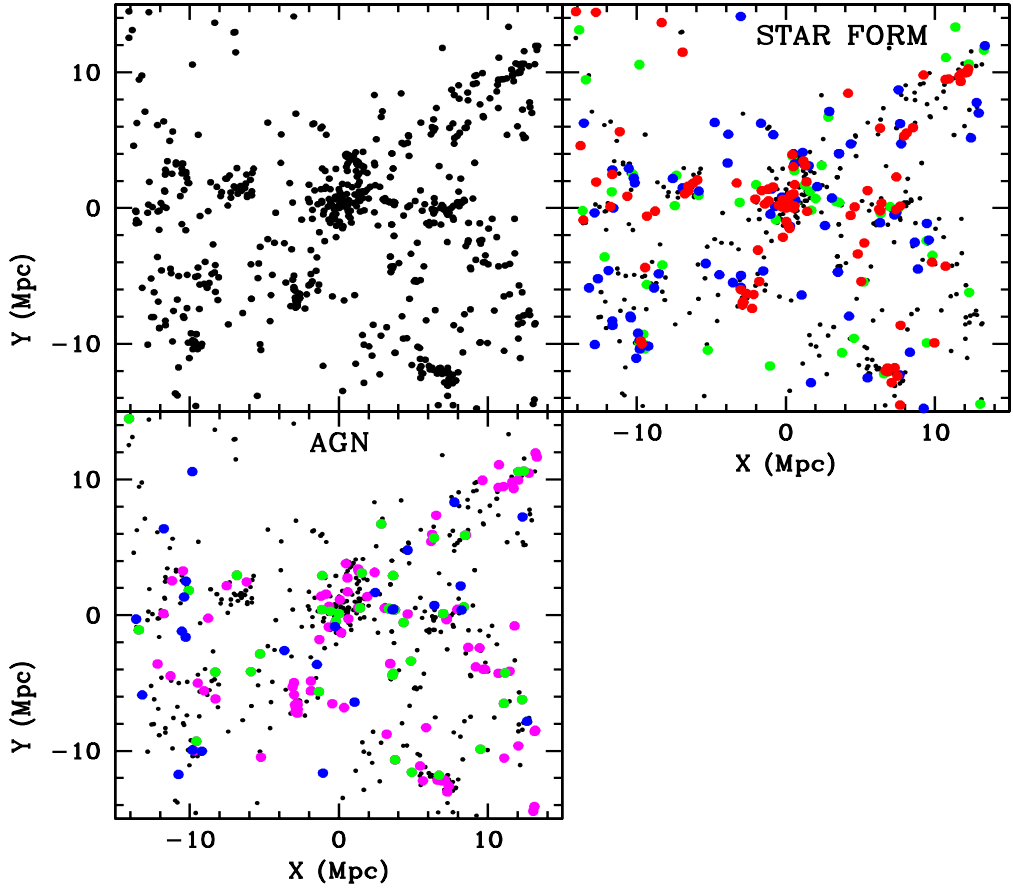}
}
\caption{\label{fig14}
\small
Top Left: The distribution of all galaxies in a `slice' at z=0.05 (see text for more details).
Top Right: Galaxies with $10^{10} M_{\odot} < M_* < 3\times 10^{10} M_{\odot}$ 
are colour-coded according
their measured 4000 \AA\ break strengths. Red is for D$_n$(4000)$>1.8$,
green is for $1.6 <$ D$_n$(4000)$<1.8$ and blue is for D$_n$(4000)$<1.6$.
(Note that galaxies falling out of the mass range are shown as
smaller black dots)
Bottom: Galaxies hosting AGN are colour-coded according to L[OIII].
Magenta is for log L[OIII]$<6.5$, green is for $6.5<$log L[OIII] $<7$,
and blue is for log L[OIII]$>7$. Galaxies with no AGN is shown as small black dots.}
\end {figure}
\normalsize

To begin, Fig. 13 illustrates how galaxies with different
star formation histories populate a typical large scale structure at z=0.05.
We searched the group and cluster catalogue of Miller et al (2004, in preparation)
for  collections of galaxy groups with very similar
redshifts spread over a field of less than 10 square
degrees. We found several such associations  
and Fig. 13 shows an example of a  rich system of galaxy groups at z=0.05. We have placed the 
origin of our x-and y--axes at the centre of the largest group in the field, which has a velocity
dispersion of $\sim 500$ km s$^{-1}$. We plot a  `slice' that includes
all galaxies with $cz$ within  500 km s$^{-1}$ of the brightest
galaxy in the group. The structures shown in Fig. 13  occupy a relatively
thin sheet that is aligned perpendicular to the line-of-sight.
As can be seen, the densest structures in the sheet appear to be arranged along
filaments that extend over scales of several tens of Mpc. 

The top left panel in Fig. 13 shows the distribution of all the galaxies in the slice  
with spectroscopic redshifts. In the top right panel we have colour-coded
galaxies with $10^{10} M_{\odot} < M_* < 3\times 10^{10} M_{\odot}$
according to their measured 4000 \AA\ break strengths. Red indicates galaxies with
D$_n$(4000)$> 1.8$, green is for galaxies with $1.6 <$ D$_n$(4000)$< 1.8$, and
blue is for galaxies with D$_n$(4000)$< 1.6$.
As shown in Paper II, field galaxies 
with $M_* < 3 \times 10^{10} M_{\odot}$ generally have D$_n$(4000) $< 1.6$. 
One should thus be able to evaluate  the scale over 
which star formation responds to local density  
simply by looking at the distribution of the                      
red and green points in the diagram.  
As can be seen, clumps of red galaxies appear wherever there is a significant
overdensity in the distribution of galaxies, even when the overdensity
only exists on small ($< 1-2$ Mpc) scales. In the bottom panel we have coloured-coded galaxies
with detected AGN according to their measured [OIII] line luminosities. The qualitative
picture is very similar.  AGN with high values of L[OIII] are found predominantly in low-density
regions, while low-luminosity AGN are also found in denser groups. 

Fig. 14 places our result on more quantitative footing. We take all the galaxies
in our sample with $0.04 < z < 0.06$ and calculate counts within a nested series of shells
in projected radius. We keep the allowed velocity difference fixed at 500 km s$^{-1}$.
Note that tracer galaxies are now defined as galaxies with $r < 17.77$ at z=0.06,
so target galaxies will have more neighbours in this analysis.   
The top right panel of Fig. 14 shows how the 
4000 \AA\ break strength of  galaxies with $10^{10} M_{\odot} < M_* < 3 \times 10^{10} M_{\odot}$
correlates with $N_{neighb}$ calculated within a projected radius of 1 Mpc.  
We have ordered all the galaxies in increasing order of  $N_{neighb}$ and we 
then divide them into groups of 150. We  plot the median value of D$_n$(4000)
in each group as a function of the median value of N$_{neighb}$ in that group.
As can be seen, there is a very tight  correlation between D$_n$(4000)
and galaxy count on scales less than 1 Mpc. In the top right panel of Fig. 14, we
plot the same thing, except that $N_{neighb}$ is evaluated within a radius of 5 Mpc.
As can be seen, D$_n$(4000) does correlate  with N$_{neighb}$ even on these
large scales, but the correlation is significantly weaker and noisier.

We can also ask  whether the density on large scales has any 
influence on the star formation history of a galaxy, once the density on small scales
is specified. The answer to this is shown in the two bottom panels of Fig. 14
where we plot D$_n$(4000) as a function of the number of neighbours with         
$1 < R_{proj} < 3$ Mpc, for galaxies with 0-1 neighbours inside 1 Mpc (bottom left)  
and for galaxies with  8-12 neighbours inside 1 Mpc (bottom right). As can be seen, the count 
within the 1-3 Mpc shell has no influence on D$_n$(4000).  
Finally, we have divided our galaxy sample into ten different bins in density
using $R_{proj} < 1$ Mpc and                                                                
we calculate the linear correlation coefficient $r$ between D$_n$(4000) and the
galaxy
count in the shell with $1 < R_{proj} < 3$ Mpc. We find values of $r$ that scatter
around zero and the distribution is entirely consistent with the null 
hypothesis of no correlation.
We note that Balogh et al (2003) carried out a similar analysis and obtained
the opposite result. We do not currently understand the reason for this.

\begin{figure}
\centerline{
\epsfxsize=13cm \epsfysize=13cm \epsfbox{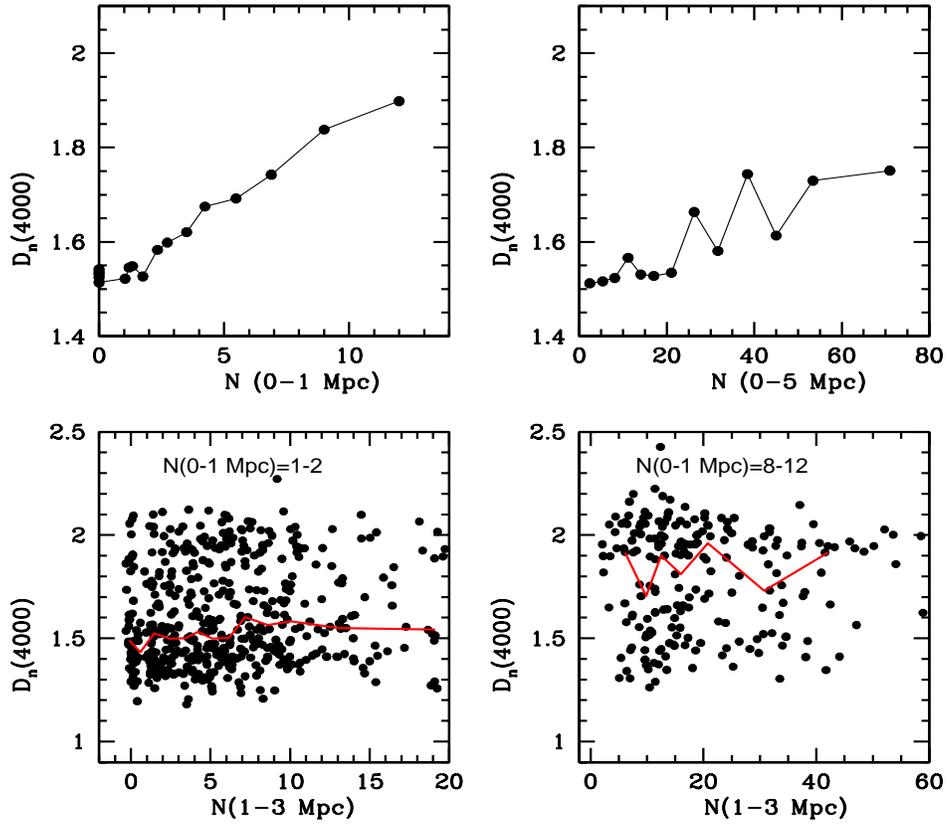}
}
\caption{\label{fig14}
\small
Top: The correlation between D$_n$(4000) and the number of neighbours within a projected radius
of 1 Mpc (left) and 5 Mpc (right).  
The solid circles indicate the median value of D$_n$(4000) at  given
value of the count.
We have chosen our bins so that they always include a fixed number  
of galaxies (150). Bottom: D$n$(4000) is plotted as a function of the count within the
shell with $1<R_{proj}< 3$ Mpc  for galaxies with a fixed value of the count within 1 Mpc.
The red line is a running median, evaluated such that fixed
number of  galaxies are included in every bin.}
\end {figure}
\normalsize

\section {Summary and Discussion}

\subsection {Summary of the Main Observational Results}

\begin{enumerate}
\item  The characteristic stellar mass of galaxies increases as a 
function of density.
\item For stellar masses above $3 \times 10^{10} M_{\odot}$, the distribution 
of size and concentration at fixed stellar mass is independent of local density.
Galaxies with $M_*< 10^{10} M_{\odot}$ tend to be slightly more
concentrated and more compact in the densest regions.
\item The galaxy property that is most sensitive to environment is 
star formation history. The relations between  D$_n$(4000) or SFR/$M_*$ and galaxy mass
depend strongly on  density. The dependence is strongest
for galaxies with stellar masses less than  $3\times 10^{10} M_{\odot}$.
For these systems, the median SFR/$M_*$ changes by more than a 
factor of 10 from our lowest density
to our highest density bin. The dependence is weaker for high
mass galaxies, but it remains significant.
\item The correlation between star formation history and local density is strongest
on small ( 1 Mpc) scales. We see no evidence that the density on large scales has any influence   
on the star formation history of a galaxy once the density on small scales is specified.
\item The relations between our three different indicators of recent star formation
history -- the 4000 \AA\ break strength  D$_n$(4000), the Balmer-absorption index H$\delta_A$,
and the normalized star formation rate SFR/$M_*$-- do not depend on environment.
\item A larger fraction of galaxies host AGN with high [OIII] line luminosities  
in low density environments. At a given value of L[OIII], the properties of
AGN hosts do not appear to depend on environment.
\item Galaxies in low-density environments contain more dust. This is true even for
the most massive galaxies in our sample with $M_* > 10^{11} M_{\odot}$.
\end {enumerate}

\subsection{From Local Density to Halo Mass}
In order to interpret our results, it is critical to understand how             
physical conditions change as a function of the particular local density 
estimates that we have used in our analysis. 

As discussed in section 3, one advantage
of characterizing the environment of a galaxy in terms of its membership in a group or cluster,
is that it is  possible to assign each object to a dark matter halo of a well-defined          
mass. Inside such a halo, galaxies move with a known                      
distribution of velocities and the thermodynamic state of the intergalactic gas is
also reasonably well-constrained. 
The link between our local galaxy density estimate and halo mass is not as straightforward,
but considerable insight may be gained by analyzing mock galaxy catalogues
generated using cosmological N-body simulations.

In this section, we make use of the publically available dark matter halo
and galaxy catalogues from the GIF simulations
described in\\ 
Kauffmann et al (1999; http://www.mpa-garching.mpg.de/galform/virgo/hrs/)
to illustrate how this may be achieved in practice. These simulations are for standard
$\Lambda$CDM initial conditions ($\Omega=0.3$, $\Lambda=0.7$, $h=0.7$, $\sigma_8=0.9$)
and were carried out in a periodic box of length 141 $h^{-1}$ Mpc. The mass of individual
dark matter particles in the simulation  is $1.4 \times 10^{10} h^{-1} M_{\odot}$.
As explained in Kauffmann et al, the history of a galaxy that forms 
inside a dark matter halo containing more than 100 particles can be tracked 
reasonably reliably in these simulations. The GIF catalogues are thus limited
to galaxies with stellar masses larger than  $2 \times 10^{10} M_{\odot}$ or
$r$-band magnitudes brighter than $\sim -20.5$. This corresponds rather well
to the magnitude limit of our volume-limited sample of SDSS tracer galaxies.

\begin{figure}
\centerline{
\epsfxsize=13cm \epsfysize=8cm  \epsfbox{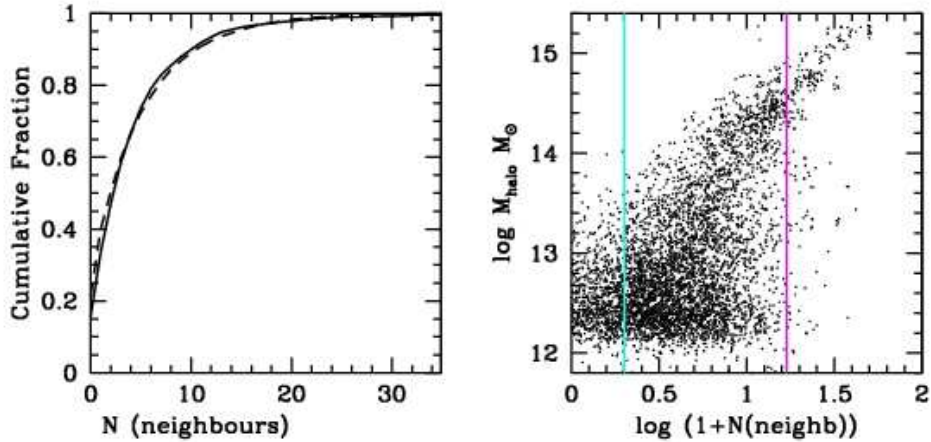}
}
\caption{\label{fig15}
\small
Left: The cumulative fraction of galaxies with counts less than $N$ for the simulated galaxies
(solid) and for the observed galaxies (dashed). 
Right: The virial mass of the halo in which a galaxy resides is plotted as a function     
of the logarithm of the counts around the galaxy. 
A random number in the range 0-0.5 has been added to each count in order to
obtain a uniform distribution in the abscissa.
The vertical  cyan and magenta lines indicate the locations of our two extreme bins in local density.}
\end {figure}
\normalsize

We have placed a cylinder of radius 2 Mpc and length 14.3 Mpc around each
galaxy in the catalog and we count the total number of galaxies contained
inside each cylinder. The left-hand panel of Fig. 15 compares the distribution of counts
around our simulated galaxies (solid line) with the  distribution of
counts around the observed galaxies  shown in Fig. 1. As can be seen, 
the two distributions agree well.     
In the right-hand panel we show a scatter-plot of  the mass of the dark matter halo in which the
simulated galaxy resides as a function of the count within the cylinder. 
This shows that galaxies with less than two neighbours 
reside in halos with virial masses
between $10^{12}$ and $10^{13} M_{\odot}$. Note that these galaxies occupy 
the density bin represented by cyan lines in the figures shown in  section 6.
In this bin, one
is thus studying a population of objects where there is typically only
one bright galaxy per halo.  Galaxies with more than 17 neighbours
(represented by magenta lines in section 6) nearly always reside in halos
more massive than $10^{14} M_{\odot}$, i.e. in galaxy clusters.
Galaxies in our intermediate density bins occupy halos with a rather wide
range in mass. However, the median halo mass increases as a a function of
the number of neighbours, which explains why galaxy properties  
change continuously as a function of our local density estimates.

We note that the above analysis is only meant to be illustrative. 
We have not yet studied whether our simulated galaxies match the observations
in detail. In addition, the GIF simulations lack the resolution to study galaxies
with  stellar masses below $\sim 10^{10} M_{\odot}$. 
However, the trend for galaxies with higher local density estimates to occupy more
massive halos should be robust.

\subsection {From Dark Matter Halos to Galaxies}

Once we make a link between the estimate of the density around a galaxy  and the  
characteristic mass of the halo in which it resides,
the observed trends 
place important constraints on galaxy formation processes.

For example, Fig. 3 shows that most of the change in the characteristic
stellar masses of galaxies occurs in our {\em low-density} (cyan and blue) bins.
The partition function of the stellar mass remains almost invariant in our
higher-density (green, black and red) bins. If we now make the identification with
halo mass shown in Fig. 15, we infer that the stellar mass of galaxies increases with halo
mass for halos with virial masses less than a few $\times 10^{13} M_{\odot}$.
In more massive halos, the distribution of stellar masses no longer depends on
$M_{halo}$. One way to explain this trend is if galaxies no longer form stars
once they fall into high mass halos. This is  qualitatively supported by
the data, which show that in general,  the star formation rates of galaxies decrease
strongly in high density environments. However, the nature of this decrease
is clearly quite complicated. Figs. 2 and 3 show that
the fraction of the total stellar mass  in star-forming galaxies decreases
continuously with density. In Fig. 7 one sees that for low mass
galaxies ($\sim 10^{10} M_{\odot}$) the drop in SFR/$M_*$ occurs primarily in
very high density environments that contain massive halos. For high mass galaxies
($\sim 10^{11} M_{\odot}$) the overall drop in SFR/$M_*$ with density is much smaller and it takes
place mainly in our lowest density bins.

There are many different physical processes that could, in principle, play a role in
quenching star formation in massive halos. Gunn \& Gott (1972) discussed how
the interstellar material in a galaxy would feel the ram pressure of the
intracluster medium as it moves through the cluster. They calculated that
for a galaxy moving at the typical velocity of 1700 km/s through the
Coma Cluster, the ISM would be stripped in one pass. Farouki \& Shapiro (1981)
used N-body simulations to show that tidal interactions between galaxies 
in clusters are effective in destroying galactic disks, a process dubbed
``harrassment'' by Moore et al (1996). In semi-analytic models of galaxy
formation, star formation in disk galaxies is maintained by  cooling and 
infall from a surrounding halo of hot gas. When a galaxy falls into 
a large halo, this diffuse halo is stripped and star formation in the
galaxy gradually declines as it uses up its cold gas 
(e.g. Kauffmann, White \& Guiderdoni 1993; Diaferio et al 2001)
Such a gas supply-driven decline in 
the star formation rates of cluster galaxies was first suggested by Larson, Tinsley
\& Caldwell (1980) and in some recent papers it 
has also been called strangulation 
(e.g. Balogh \& Morris 2000; Balogh, Navarro \& Morris 2000).
Other ICM-driven processes that have been discussed
in the literature include evaporation (Cowie \& Songaila 1977) and turbulent viscous
stripping (Nulsen 1982).

\begin{figure}
\centerline{
\epsfxsize=14cm \epsfysize=14cm \epsfbox{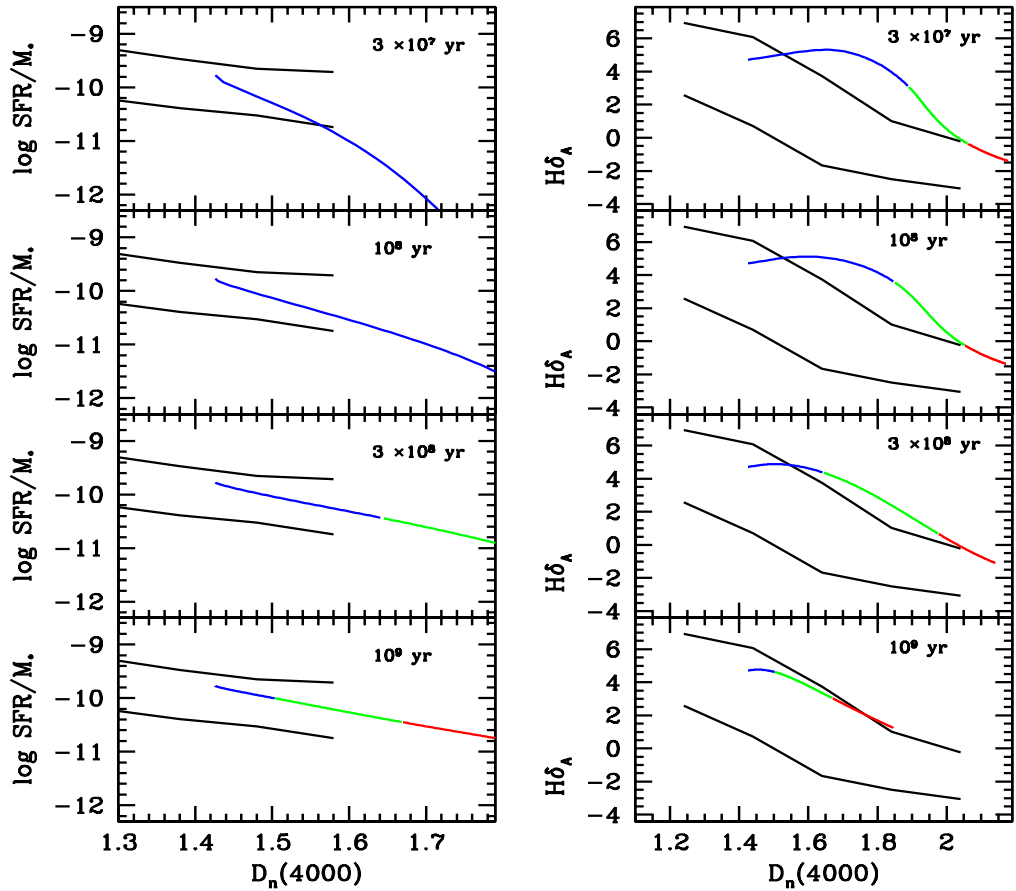}
}
\caption{\label{fig16}
\small
The evolution of a galaxy that has been forming stars at a continuous
rate for 10 Gyr following trucation in its star formation rate.
Results are shown for different quenching timescales, as defined in the text.
The blue part of curve shows the evolution for the first 0.5 Gyr following
the quenching of the star formation, green is for t=0.5-1.5 Gyr and red is for
t=1.5-2.5 Gyr. The black lines delineate the 10 and 90 percentiles
of the observed relations shown in Fig. 8. }
\end {figure}
\normalsize

We note that for massive galaxies,  harrassment is not favoured by the observations, because
it is difficult to explain why the sizes and concentrations  of galaxies
of fixed mass do not depend on environment, if disks are destroyed in high density regions.
Another way to distinguish between these different processes is through constraints
on the {\em timescale} over which the star formation is quenched in
high mass halos. If ram-pressure stripping of the interstellar medium
is the most important process,
then this timescale is expected to be short ($\sim 10^7$ years). If instead gas-supply
driven processes are the dominant factor, then the decline should be more gradual.

The facts that star-forming galaxies are currently present in low-density environments 
and that the fraction of such objects declines 
continuously with increasing density, imply that quenching processes
are still operating.
In Fig. 8, we demonstrated that the relations between three different indicators of recent star formation
history -- the 4000 \AA\ break strength, the Balmer-absorption index H$\delta_A$
and the normalized star formation rate SFR/$M_*$-- do not depend on environment.
This means that when star formation in a galaxy is quenched, its values of
D$_n$(4000), H$\delta_A$ and SFR/$M_*$ must change {\em in such a way that the galaxy evolves
along these relations}. In Fig. 16 we demonstrate that this requires  star formation
to decline on relatively long ($> 1$ Gyr) timescales.

Let us consider a galaxy that has been forming stars at a continuous rate (SF$_0$) for 10 Gyr    
(i.e  $\log \rm{SFR/M}_* \sim -10$). As shown by Brinchmann et al (2003), the 
distribution of SFR/$M_*$ for star-forming galaxies peaks at values close
to this over nearly 3 orders of magnitude in stellar mass, declining
only at $M_* >$ few $ \times  10^10 M_{\odot}$. 
According to the population synthesis models of Bruzual \& Charlot (2003),
such a  galaxy is predicted to have D$_n$(4000)$\sim 1.4$ and H$\delta_A \sim 4$.
After 10 Gyr of continuous star formation, we let the star formation rate evolve
as SFR(t)= SF$_0$ $\exp (-t/\tau_q)$, where $\tau_q$ is the characteristic timescale over
which the star formation is quenched in the galaxy. Fig. 16 shows how the galaxy moves  
in the SFR/M$_*$-D$_n$(4000) and H$\delta_A$-D$_n$(4000) planes for different values of $\tau_q$
The blue part of the curve shows the evolution for the first 0.5 Gyr after the star 
formation rate begins to decline. The green part of the curve is for t=0.5-1.5 Gyr
and red is for t=1.5-2.5 Gyr. As can be seen, galaxies move along the relations
defined by observed galaxies only if the quenching timescale is long ($\tau_q >$ 1 Gyr).
If the quenching timescale is short, the galaxy will move away from the main body
of the data and stay there for a period of up to 1.5 Gyr. The H$\delta_A$-D$_n$(4000)
relation turns out to be more constraining than SFR/M$_*$-D$_n$(4000), 
because it is defined for all
the galaxies in the sample, not just for those with emission lines and young stellar ages.

Although we have argued star formation in the {\em majority} of galaxies is being
quenched over long timescales, clear exceptions do exist. 
\hspace {0.1cm} Quintero et al (2003)
have identified a population of ``K+A'' galaxies in the SDSS with spectra
exhibiting strong Balmer absorption lines, but with little or no
H$\alpha$ emission. However, such galaxies are currently very rare ($<$1\% of the
total population). 
Balogh et al (2003) have  argued that the timescale for
quenching processes ought to be {\em short}, citing the fact that 
the  distributions of H$\alpha$ equivalent widths of star forming galaxies
are independent of environment. Our arguments contradict this.
Fig. 16 shows that for slow quenching galaxies should move along the
observed relations, while rapid quenching would lead to
an overabundance of ``K+As''.  More detailed modelling
is clearly necessary in order to settle this question.                   
An  estimate of the recent accretion rate
of galaxies onto massive halos is required
in order to set tight limits on  quenching timescales. In addition, it is necessary to adopt
a range of different star formation histories  in order
to account for the observed scatter of our `blue'  galaxies in the H$\delta_A$-D$_n$(4000) plane.
We will leave more detailed analysis of this kind for a future paper.

\subsection {From Low Density to High Redshift}

In section 7.1, we showed that there is a correspondence between our estimate of the local  
density around a galaxy and the mass of the dark matter halo in which it resides.    
In the now-standard $\Lambda$CDM cosmology, structure in the Universe builds up via a process
of hierarchical clustering, with small structures merging together to form
larger and larger objects. At high redshifts,  the abundance of  massive halos decreases and fewer
galaxies reside in clusters and rich groups. This is illustrated in Fig. 17 where we show how the
dark matter is partitioned among halos of different masses in
our $\Lambda$CDM simulation at a series of different redshifts.  The fraction of the matter 
that has collapsed and fallen into halos with masses less than $10^{13} M_{\odot}$  remains
approximately constant from $z=0$ to $z=1$. However,  the fraction of material in  halos
with masses greater than $10^{14} M_{\odot}$ decreases from 12 percent  at $z=0$ to 1.5 percent at z=1.
By redshift 2, the fraction of matter in halos between $10^{12}$ and $10^{13} M_{\odot}$
has also begun to decrease.

\begin{figure}
\centerline{
\epsfxsize=8cm \epsfysize=8cm \epsfbox{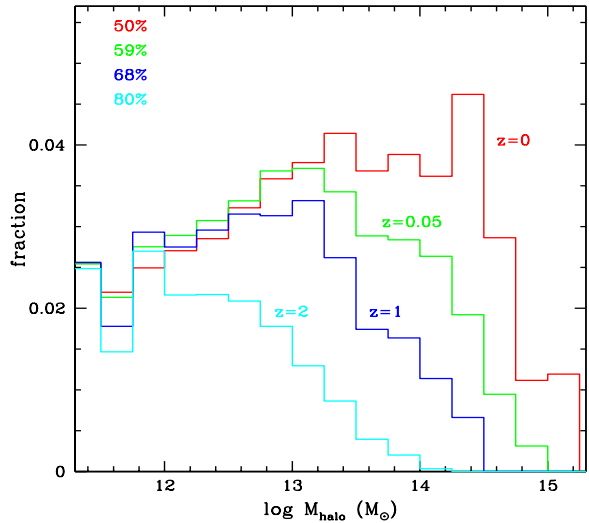}
}
\caption{\label{fig17}
\small
The fraction of the dark matter in the Universe in halos of different mass. }
\end {figure}
\normalsize

In one  compares Fig. 17 with Fig. 15, one might naively speculate that `field' galaxies at $z \sim 1$ would
look much like galaxies in our low-density (cyan and blue) bins. Obviously the real situation is     
complicated by the fact that the Universe was only half its present age at $z=1$.  Galaxies 
were therefore less `evolved' than at the present day; they were probably  more gas-rich and 
 their stellar populations were younger and hence bluer
and brighter. Nevertheless when one examines the trends in our data, a  {\em qualitiative analogy}
between galaxies in low-density environments and galaxies in the high-redshift Universe seems quite
compelling.

The most striking evolutionary trend in the galaxy population from $z=0$  to $z \sim 1-2$
is the rapid increase in the global star formation rate density (Lilly et al 1996 ; Madau et al 1996).
The reason for this evolution  was first pointed out by Cowie et al (1996), who showed that
the maximum rest-frame K-luminosity of galaxies that undergo rapid star formation
has been declining smoothly with redshift from $z \sim 1$ to the present. Cowie et al. dubbed
this process `down-sizing' and this is exactly what is seen in Fig. 7. As local density
increases, strong star formation occurs only in galaxies with progressively smaller and smaller stellar
masses.

There has been considerable recent attention has focused on a population of extremely red objects (EROs) at
$z >  1$ (e.g. Elston et al 1988; Daddi, Cimatti \& Renzini  2000; Cimatti et al 2002). 
These are massive galaxies with colours
as red as (or in some cases even redder than) those of passively evolving elliptical
galaxies. When studied spectroscopically, some of these galaxies  turn out
to be genuinely old systems, but a substantial fraction are forming stars and are red because they
are very dusty. In fig. 12 we showed that as local density decreases, there are a larger
fraction of  very massive
galaxies with star formation and strong dust attenuation.

Finally, it is interesting that  quantities that are only weakly dependent 
on local density also seem to exhibit little change with redshift. Even though the inferred
characteristic  mass of dark matter halos  decreases by a factor of $\sim 100$ from our highest
density to our lowest density bins, Fig. 2 shows that the
characteristic stellar masses of the galaxies in these halos decreases by only a factor                       
of $ \sim 2$. 
Studies of the evolution of the rest-frame
near-IR luminosity function suggest that characteristic stellar masses of galaxies have
evolved only weakly from $z \sim 1$ to the 
present (e.g. Pozzetti et al 2003; Somerville et al 2003).
In Fig. 6, we showed that at given stellar mass, the sizes and concentrations of galaxies
exhibit almost no dependence on environment. In a recent study of 168 galaxies with K-band
magnitudes brighter than 23.5, Trujillo et al (2003) find that the relation between 
galaxy size and stellar mass has remained unchanged since $z \sim 3$. A study of the size-mass
relation of several tens of  thousands of galaxies  in the GEMS/COMBO-17 survey leads to
the same basic conclusion (MacIntosh et al, in preparation; Barden et al, in preparation). 

\subsection {Galaxy Formation: Nature or Nurture?}

We began this paper by asking whether the correlations
between the different physical properties of galaxies were imposed 
very early on (the ``nature''
hypothesis) or whether they are the end product of processes that have operated  
over a Hubble time (the ``nurture'' scenario).
Our tentative, perhaps unsurprising conclusion,  
is that both nature and nurture are important for
understanding the evolution of the galaxy population. 

We have shown that the relations
between between galaxy structural parameters and stellar mass exhibit
little dependence on density. This suggests that these relations were
in place relatively early on. As discussed above, studies of the distribution of galaxy sizes
at high redshift also seem to support this conclusion.

We have also shown that the relation between star formation history and stellar mass
is quite sensitive to local density. We note that most of the density-dependence
is for galaxies with $ M_* \sim 10^{10} M_{\odot}$. Very massive galaxies generally have old
stellar populations, both in low density and in high density environments.
Very low mass galaxies 
tend to have young stellar  populations and exhibit only a weak drop in
average star formation rate from our lowest density to our highest density bins.

Fig. 18 attempts to put these two findings together.                                   
In the top panels we plot D$_n$(4000) as a function of stellar mass
for galaxies in our lowest density bin. Results are
shown separately for disk-dominated galaxies with $C< 2.6$ (left) and for
bulge-dominated galaxies with $C>2.6$ (right). We only
plot galaxies with $z < 0.055$ so that our
sample is  volume-limited down to the lowest stellar masses shown on the plot.
As can be seen, the division
of the galaxy population into two distinct ``families'' (old, massive and
concentrated; small, young and diffuse) persists even in our lowest
density bin.  In the bottom panel, we show the same for galaxies
in our two highest density bins. We see  that the ``tail''
of massive, star-forming early-type galaxies present in
the low density bin  has diminished and has been
replaced by  a uniform population of old ``ellipticals''. 
At the same time, a significant number of disk-dominated galaxies with
lower masses have stopped forming stars and have moved onto the red
sequence. 

\begin{figure}
\centerline{
\epsfxsize=14cm \epsfysize=14cm \epsfbox{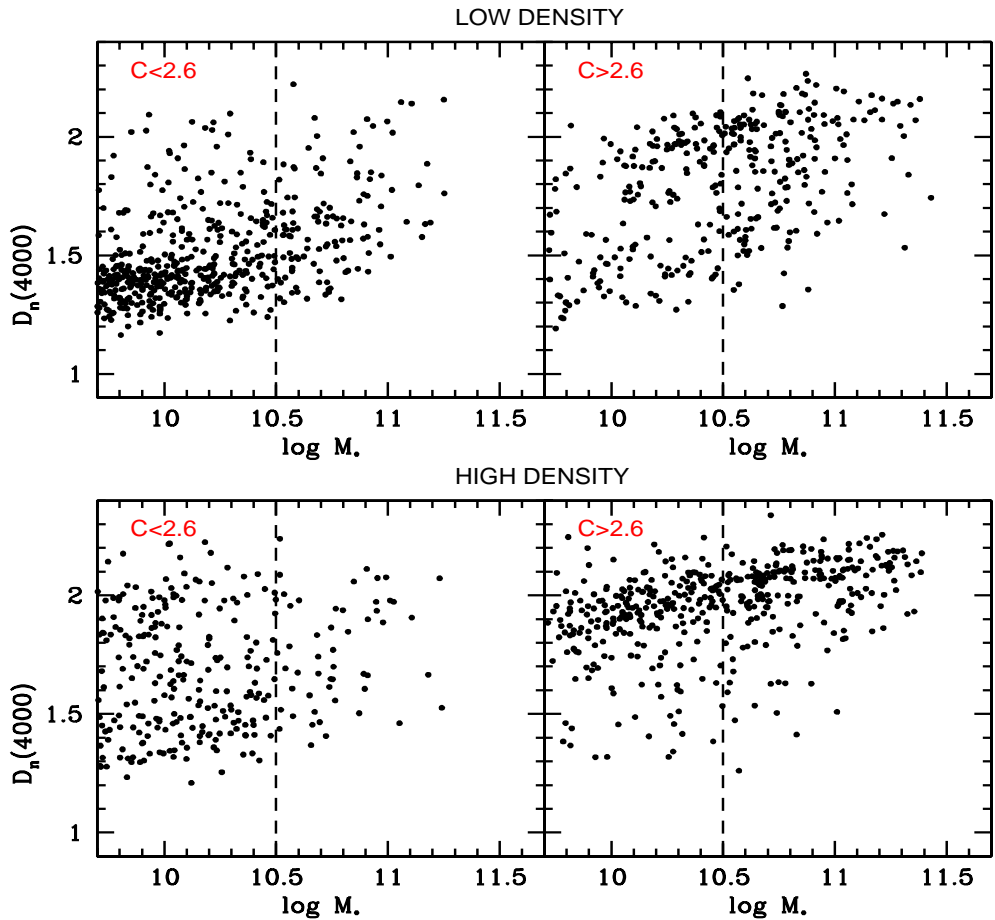}
}
\caption{\label{fig18}
\small
Top: D$_n$(4000) is plotted as a function of stellar mass for galaxies with
$C<2.6$ (left) and $C>2.6$ in our lowest density bin (cyan in Fig. 2). 
Bottom: The same, except for galaxies in our two  highest density bins
(black and red in Fig. 2).
We have plotted galaxies with $z<0.055$ so that the samples are volume-limited
down to the lowest stellar masses shown in the plot.} 
\end {figure}
\normalsize
Our results suggest that there are two basic routes by which galaxies 
reach the end-point of the evolution and become ``dead and red''. 
The first is an event, for example a merger, that leads to the formation of
a massive bulge and an associated black hole. During the event stars form very rapidly, 
the internal gas reservoir of the galaxy is depleted and subsequent
cooling is strongly suppressed.
The second route is the removal of the cold gas supply   
by {\em external processes} 
that operate in  massive halos. This has very little effect on the structure of massive
galaxies, which already have low gas fractions.

The majority of massive galaxies passed along the first route at relatively
high redshifts and now have high concentrations and surface mass densities.
Low mass galaxies, on the other hand, generally do not pass along this route
and in low-density environments  
they are still actively forming stars. As low
mass galaxies fall into massive halos, their cold gas supply is removed 
and their star formation shuts down.
Massive galaxies are {\em already} near the
end-point of their evolution when they fall into massive
halos, but the same external process
at work on the low-mass galaxies finishes the job. This not only
shuts off any residual star formation in these systems, but also starves
the black hole.

Considerable work remains to be done in understanding the {\em physics} that 
leads to the scaling relations
presented in this paper. We believe that a combination of data 
from the new generation of large surveys,
both at $z=0$ and at high redshifts, with insight 
gained from modern simulations of structure formation
will be the key to unravelling how galaxies formed and evolved.

\vspace{1.0 cm}
S.C. thanks the Alexander von Humboldt Foundation, the Federal
Ministry of Education and Research, and the Programme for Investment
in the Future (ZIP) of the German Government for their support.
B.M. thanks the Florence Gould Foundation.

Funding for the creation and distribution of the SDSS Archive has been provided
by the Alfred P. Sloan Foundation, the Participating Institutions, the National
Aeronautics and Space Administration, the National Science Foundation, the U.S.
Department of Energy, the Japanese Monbukagakusho, and the Max Planck Society.
The SDSS Web site is http://www.sdss.org/.

The SDSS is managed by the Astrophysical Research Consortium (ARC) for the
Participating Institutions. The Participating Institutions are The University
of Chicago, Fermilab, the Institute for Advanced Study, the Japan Participation
Group, The Johns Hopkins University, Los Alamos National Laboratory, the
Max-Planck-Institute for Astronomy (MPIA), the Max-Planck-Institute for
Astrophysics (MPA), New Mexico State University, University of Pittsburgh,
Princeton University, the United States Naval Observatory, and the University
of Washington.

\pagebreak 
\Large
\begin {center} {\bf References} \\
\end {center}
\normalsize
\parindent -7mm  
\parskip 3mm

Abazaijan, K. et al, 2003, AJ, 126, 2081

Baldry, I.K., Glazebrook, K., Brinkmann, J., Ivezic, Z., Lupton, R.H., Nichol, R.C.,
Szalay, A.S, ApJ, in press (astro-ph/0309710) 

Balogh, M.L., Morris, S.L., 2000, MNRAS, 318, 703

Balogh, M.L., Navarro, J.F., Morris, S.L., 2000, ApJ, 540, 113

Balogh, M.L., Christlein, D., Zabludoff, A.I., Zaritsky, D., 2001, ApJ, 557, 117

Balogh, M.L. et al, 2003, submitted (astro-ph/0311379)

Benson, A.J., Cole, S., Frenk, C.S., Baugh, C.M., Lacey, C., 2000, MNRAS, 311, 793

Birnboim, D., Dekel, A., 2003, MNRAS, 345, 349

Blanton, M.R. et al, 2003a, ApJ, 594, 186 

Blanton, M.R., Eisenstein, D.J., Hogg, D.W., Schlegel, D.J., Brinkmann, J, 2003b, ApJ, submitted 
(astro-ph/0310453)

Blanton, M.R. et al, 2003c, AJ, 125, 2348 

Blanton, M.R., Lupton, R.H., Maley, F.M., Young, N., Zehavi, I., 
Loveday, J. 2003d, AJ, 125, 2276

Bower, R.G., 1991, MNRAS, 248, 332

Brinchmann, J., Charlot, S., White, S.D.M., Tremonti, C., Kauffmann, G., Heckman, T.M.,
Brinkmann, J., 2003, MNRAS, submitted (astro-ph/0311060)

Bruzual, G., Charlot, S., 2003, MNRAS, 344, 1000

Budavari, T., et al, 2003, ApJ, 595, 59 

Charlot, S., Longhetti, M., 2001, MNRAS, 323, 887

Cimatti, A. et al, 2002, A\&A, 381, L68

Daddi, E., Cimatti, A., Renzini, A., 2000, A\&A, 362, L45

Davis, M., Geller, M.J., 1976, ApJ, 208, 13

Diaferio, A., Kauffmann, G., Balogh, M.L., White, S.D.M., Schade, D., Ellingson, E., 2001,
MNRAS, 232, 999

Cowie, L.L., Songaila, A., 1977, Nature, 266, 501

Cowie, L.L., Songaila, A., Hu, E.M., Cohen, J.G., 1996, AJ, 112, 839

De Propris, R. et al, 2003, MNRAS, 342, 725

Diaferio, A., Kauffmann, G., Balogh, M.L., White, S.D.M., Schade, D., Ellingson, E., 2001, MNRAS, 328, 726

Dressler, A., 1980, ApJ, 236, 531

Elston, R., Rieke, G.H., Rieke, M.J., 1988, ApJ, 331, L77

Farouki, R., Shapiro, S.L., 1981, ApJ, 243, 32

Fukugita, M., Ichikawa, T., Gunn, J.E., Doi, M., Shimasaku, K., 
Schneider, D.P. 1996, AJ, 111, 1748

Gomez, P.L. et al, 2003, ApJ, 584, 210

Goto, T., Yamaguchi, C., Fujita, Y., Okamura, S., Sekiguchi, M., Smail, I.,
Bernardi, M., Gomez, P., 2003a, MNRAS, 346, 601

Goto, T. et al, 2003b, PASJ, 55, 757

Gunn, J., Carr, M., Rockosi, C., Sekiguchi, M., Berry, K., Elms,
B., de Haas, E., Ivezi\'{c}, Z. et al. 1998, ApJ, 116, 3040

Hashimoto,Y., Oemler, A., Lin, H., Tucker, D.L., 1998, ApJ, 499, 589 

Hogg, D., Finkbeiner, D., Schlegel, D., \& Gunn, J. 2001, AJ, 122, 2129

Hogg, D.W. et al, 2003a, ApJ, 585, L5

Hogg, D.W. et al, 2003b, ApJ, submitted (astro-ph/0307336)

Hoyle, F., Rojas, R.R., Vogeley, M.S., Brinkmann, J., ApJ, submitted (astro-ph/0309728)

Kauffmann, G., White, S.D.M., Guiderdoni, B., 1993, MNRAS, 264, 201

Kauffmann, G., Colberg, J.M., Diaferio, A., White, S.D.M., 1999, MNRAS, 303, 188

Kauffmann, G. et al, 2003a, MNRAS, 341, 33 (Paper I) 

Kauffmann, G. et al, 2003b, MNRAS, 341, 54 (Paper II) 

Kauffmann, G. et al 2003c, MNRAS, 346, 1055                         

Kodama, T., Smail, I., Nakata, F., Okamura, S., Bower, R.G., 2001, ApJ, 562, L9

Koopmann, R.A., Kenney, J.D.P., 1998, ApJ, 497, L75

Lacey, C., Cole, S., 1993, MNRAS, 262, 627

Larson, R.B., Tinsley, B.M., Caldwell, C.N., 1980, ApJ, 237, 692 

Lemson, G., Kauffmann, G., 1999, MNRAS, 302, 111

Lewis, I., et al, 2003, MNRAS, 334, 673

Lilly, S.J., Le Fevre, O., Hammer, F., Crampton, D., 1996, ApJ, 460, L1

Madau, P., Ferguson, H.C., Dickinson, M.E., Giavalisco, M., Steidel, C.C., Fruchter, A., 1996,
MNRAS, 283, 1388

Loveday, J., Maddox, S.J., Efstathiou, G., Peterson, B.A., 1995, ApJ, 442, 457

Madgwick, D.S. et al, 2003, MNRAS, 344, 847

Marinoni, C., Davis, M., Newman, J.A., Coil, A.L., 2002, ApJ, 580, 122

Miller, C.J., Nichol, R.C., Gomez, P.L., Hopkins, A.M., Bernardi, M., 2003, ApJ, 597, 142 

Moore, B., Katz, N., Lake, G., Dressler, A., Oemler, A., 1996, Nature, 379, 613

Nolthenius, R., White, S.D.M., 1987, MNRAS, 225, 505

Norberg, P. et al , 2001, MNRAS, 328, 64

Nulsen, P.E.J., 1982, MNRAS, 198, 1007

Oemler, A., 1974, ApJ, 194, 1

Pier, J.R., Munn, J.A., Hindsley, R.B., Hennessy, G.S., Kent, S.M.,
Lupton, R.H., Ivezi\'{c}, Z., 2003, AJ, 125, 1559

Pimbblet, K.A., Smail, I., Kodama, T., Couch, W.J., Edge, A.C., Zabludoff, A.I., 
O'Hely, E., 2002, MNRAS, 331, 333

Pozzetti, L et al, 2003, A\&A, 402, 837

Quintero, A.D., et al, 2004, ApJ, submitted (astro-ph/0307074)

Richstone, D.O., 1976, ApJ, 204, 642

Smith, J.A., et al 2002, AJ, 123, 2121

Somerville, R.S. et al, 2003, ApJ, in press (astro-ph/0309067)

Shimasaku, K. et al, 2001, AJ, 122, 1238                                                

Spergel, D.N. et al, 2003, ApJS, 148, 175

Stoughton, C. et al, 2002, AJ, 123, 485

Strateva, I. et al, 2001, AJ, 122, 1861                                               

Strauss, M., et al 2002, AJ, 124, 1810  

Tanaka, M. et al, 2003, in preparation

Toomre, A., Toomre, J., 1972, ApJ, 178, 623

Tremonti, C.A. et al, 2004, ApJ, submitted

Trujillo, I. et al, 2003, ApJ, submitted (astro-ph/0307015)

Van den Bergh, S., 1976, ApJ, 206, 883

White, S.D.M., Frenk, C.S., 1991, ApJ, 379, 52

Willmer, C.N.A., Da Costa, L.N., Pellegrini, P.S., 1998, AJ, 115, 869

York D.G. et al, 2000, AJ, 120, 1579                                                             

Zehavi, I. et al., 2002, ApJ, 571, 172

\end{document}